\documentclass[prd,twocolumn]{revtex4}
\usepackage{dcolumn}% Align table columns on decimal point
\usepackage{multirow}% Columns spanning multiple rows
\usepackage{amsmath,amsthm,amssymb}
\usepackage{hyperref}
\usepackage{xcolor}

\def\fsc{A}
\def\lsc{\mathpzc{l}}

\def\Nsc{\mathpzc{N}}
\def\Psc{\mathpzc{P}}

\newcommand{\pert}[2]{
{}^{(#1)}\hspace{-0.5mm}#2
}

\DeclareMathAlphabet{\mathpzc}{OT1}{pzc}{m}{it}

\begin{document}
%

%\rightline{CERN-TH-2017-023}
%\vskip 3mm

%\preprint{CERN-TH-2017-023}
\title{
Towards a theory of nonlinear gravitational waves:
\\
a systematic approach to nonlinear gravitational perturbations in vacuum
}

\author{Andrzej Rostworowski}
\email{arostwor@th.if.uj.edu.pl}
\affiliation{M. Smoluchowski Institute of Physics, Jagiellonian University, 30-348 Krak\'ow, Poland}
%\affiliation{Theoretical Physics Department, CERN, CH-1211 Geneva 23, Switzerland}
%
%\date{\today}
%
\begin{abstract}
We present a systematic and robust approach to nonlinear gravitational perturbations of vacuum spacetimes. This approach provides a basis for \textit{a theory of nonlinear gravitational waves}. In particular, we show that the system of perturbative Einstein equations reduces at each perturbation order to two (for each gravitational mode in $3+1$ dimensions on which our study is focused) scalar wave equations, and then we show how the metric perturbations can be explicitly obtained, once the solutions to these scalar wave equations are known. These results show that the concept of polarization of a gravitational wave does make sense also beyond the linear approximation.
\end{abstract}

\maketitle

%%%%%%%%%%%%%%%%%%%%%%%%%%%%%%%%%%%%%%%%%%%%%%%%%%%%%%%%%%%%%%%%%%%%%%%%%%%%%%
%%%%%%%%%%%%%%%%%%%%%%%%%%%%%%%%%%%%%%%%%%%%%%%%%%%%%%%%%%%%%%%%%%%%%%%%%%%%%%
%%%%%%%%%%%%%%%%%%%%%%%%%%%%%%%%%%%%%%%%%%%%%%%%%%%%%%%%%%%%%%%%%%%%%%%%%%%%%%
\section{Introduction.}
\label{Sec:Intro}
%%%%%%%%%%%%%%%%%%%%%%%%%%%%%%%%%%%%%%%%%%%%%%%%%%%%%%%%%%%%%%%%%%%%%%%%%%%%%%
Due to the complexity and nonlinear nature of Einstein equations, their exact, analytic solutions are few in number. Therefore to get quantitative insight in physically interesting gravitational phenomena, in particular in a dynamical setting, one has to resort either to (complex) numerical simulations or to perturbative methods. Perturbations of exact solutions have been studied in the context of gravitational radiation and black holes stability \cite{RW, Zerilli}, self-force, accretion disks around black holes, extreme-mass-ratio inspirals and cosmology. Although there is a vast literature on linear perturbations, there are only few papers (as far as we know) dealing with second order gravitational perturbations (and they are mainly devoted to second order perturbations of Schwarzschild black hole, see \cite{GNPP,GP,NI,BMGT} and references therein; nonlinear gravitational perturbations of anti--de Sitter space were studied for the first time in \cite{DHS}, in the approach based on \cite{KI2}). Moreover, nonlinear perturbations are usually treated case by case, and a systematic approach, along some general principles, to nonlinear gravitational perturbations seems to be missing. The aim of this work is to provide a systematic and robust scheme to deal with nonlinear gravitational perturbations, in principle at any order of perturbation expansion 
\footnote{In fact, it may well happen that for many aspects of gravity the third order perturbations suitably re-summed in the form of some kinetic theory may be crucial.}. 
This scheme has been tested to work for perturbations of any spherically symmetric vacuum solutions but, in principle, it could be also applied to the Kerr solution (at the price of mode coupling already at linear order). The main idea behind our scheme is in fact extremely simple: in spite of huge complexity of gravitational perturbations, in $3+1$ dimensions, there are only two polarization states in gravitational waves! Therefore, the whole gravitational sector of perturbations should be governed at each order of perturbation expansion by only two scalar degrees of freedom (for a given radiation mode). For linear perturbations of Schwarzschild black hole these are the well known Regge-Wheeler (RW) \cite{RW} and Zerilli \cite{Zerilli} master scalar variables. 
%%%%%%%%%%%%%%%%%%%%%%%%%%%%%%%%%%%%%%%%%%%%%%%%%%%%%%%%%%%%%%%%%%%%
In this work we show how to obtain the analogues of these master scalar variables and inhomogeneous wave equations they satisfy at any order of perturbation expansion around any spherically symmetric solution of vacuum Einstein equations. Then we show how the metric perturbations can be explicitly obtained, once the solutions to these scalar wave equations are known. 
%%%%%%%%%%%%%%%%%%%%%%%%%%%%%%%%%%%%%%%%%%%%%%%%%%%%%%%%%%%%%%%%%%%%
To set the stage for the forthcoming discussion we assume that the reader is familiar with the Regge-Wheeler decomposition and gauge choices \cite{RW}; a pedagogical expositions of this subject is given in Sec.2 of excellent Nollert's review paper on quasinormal modes \cite{Nollert}.  On the basis of our recent experience, gained in studies of nonlinear gravitational perturbations of anti-de Sitter space (AdS) \cite{r2017}, we propose the following recipe to identify the master scalar variables underlying the system of perturbative Einstein equations:
%\enlargethispage{10mm}
\begin{enumerate}
\item %1) 
We assume that at each perturbation order there exist two (for each gravitational mode) master scalar variables satisfying an (inhomogeneous) scalar linear wave equation on the zero-order metric background with a (uniquely defined) potential (for the perturbations of Schwarzschild black hole these are the Regge-Wheeler \cite{RW} and Zerilli \cite{Zerilli} potentials respectively). As these master scalar variables correspond to two polarization states of a gravitational wave, this assumption simply means that the concept of polarization of the gravitational wave can be extended beyond linear approximation.
\item %2) 
At each perturbation order, all six (in 3+1 dimensions) Regge-Wheeler type gauge invariant variables are given as linear combinations of master scalar variables and their derivatives up to the second order (and also some source functions at nonlinear orders); the coefficients of these linear combinations (and the source functions at nonlinear orders) are uniquely determined by fulfilling perturbative Einstein equations, under the condition that the inhomogeneous wave equation for master scalar variables is satisfied. Thus, the RW variables are expressed in terms of master scalar variables.
\item %3)
These relations can be inverted for the master scalar variables to be given in terms of the RW variables. There are many choices that can be made here, however there is a unique choice (that corresponds to the choice made by Moncrief \cite{Moncrief} in the Schwarzschild case) such that the initial data for the wave equations for scalar master variables are given in terms of first and second fundamental forms of initial data hypersurface.
\end{enumerate}

In the next sections we will illustrate these concepts with two non-trivial examples (that together with the results of \cite{r2017} form three illustrative examples, with increasing level of complexity). Before doing it we still want to make a few general remarks.
\begin{enumerate}
\item
%\\
%\textit{Remark 1.} 
As soon as one goes beyond the linear order of perturbation expansion, gauge issues become a nuisance \cite{BMMS} and in the case of (asymptotically flat) Schwarzschild black hole a lot of effort was put into providing fully gauge invariant formulation of second order perturbations \cite{GP,BMGT}. This is not very practical at higher order calculations, and moreover, to extract radiation from the system one usually goes to the asymptotically flat gauge anyway (cf. Sec.V in \cite{GP} and Sec.IV in \cite{BMGT}). Moreover, in the fully gauge invariant formalism, the sources in the obtained scalar wave equations are often ill behaved asymptotically and there is a need for tedious and obscure procedure of regularizing these scalar sources (cf. \cite{GP,BMGT}). Thus, what we find optimal is to first solve perturbative Einstein equations in some uniquely defined gauge (we stick to the RW choice \cite{RW}) and then, before going to the next order, fix the gauge to the asymptotically desired form \cite{r2017}. In this way, the the obtained scalar sources are well-behaved asymptotically and there is no need for their regularization.
\item 
%\\
%\textit{Remark 2.} 
It is crucial for the consistency of perturbation expansion at nonlinear orders that the sources in perturbative Einstein equations (see eq. (\ref{eq:pertEq}) below) are not independent but fulfill four identities. These are the consequences of contracted Bianchi identities and are straightforward to obtain (see eqs. (\ref{eq:zero0}-\ref{eq:zero2}) and eq. (\ref{eq:zero0comoving}-\ref{eq:zero2comoving}) below). 
\item
%\\
%\textit{Remark 3.} 
After RW \cite{RW} we expand all tensors into multipoles (see (\ref{eq:polarT_ab}-\ref{eq:axialV_3}) below), however one has to remember that $\ell=0,1$ multipoles need a special treatment at nonlinear orders. At linear order, the $\ell=0,1$ multipoles can be either put identically to zero with a suitable gauge choice (if the zero order solution is maximally symmetric) or they would correspond to a shift of parameters of the zero order solution (like the mass parameter in Schwarzschild like solutions) and thus they do not contain any physically relevant information. This is no longer the case at nonlinear orders where these multipoles have to be properly included into the perturbation scheme \cite{r2017}.
\item
%\\
%\textit{Remark 4.} 
%\nopagebreak 
For pedagogical reasons we limit ourselves to axial symmetry - introducing azimuthal angle dependence (in 3+1 dimensions) is a technical, not a conceptual issue. Going beyond axial symmetry would obscure the whole discussion adding little to the points we wish to highlight in this work. Axial symmetry is also convenient, as it allows us to stay in the sector of polar perturbations only (if we start with only polar perturbations at a linear order). The approach to axial perturbations goes along the same lines as for polar perturbations, and is in fact much simpler at technical level, thus good understanding of higher order perturbations in the model case of polar perturbations at axial symmetry is sufficient to deal with more complicated cases.
\item
%\\
%\textit{Remark 5.} 
Also, for pedagogical reasons, we decided to present our approach on concrete examples, in the fixed coordinate systems (Schwarzschild -(dS/AdS) in Schwarzschild static coordinates (\ref{eq:line_element_static}) and de Sitter space in comoving coordinates (\ref{eq:line_element_comoving})) rather then discussing coordinate independent formulation akin to the formulation of Mukohyama \cite{Mukohyama} and Kodama-Ishibashi \cite{KI} at linear order of perturbation expansion. We believe that good understanding of these two examples will allow the reader to adapt our approach to the cases of his/her own interest.  
\item
%\\
%\textit{Remark 6.} 
We postpone including matter into our scheme
%, first starting from the linear order to study self-force type problems, and then also starting from the zero-order solution to study cosmological perturbations, 
to the future work. 
\end{enumerate}
The paper is organized as follows. In Sec.~\ref{Sec:general} we review from \cite{r2017} the general formulation of perturbative Einstein equations. In Secs.~\ref{Sec:Schwarzschild},\ref{Sec:deSitter} we discuss our two examples: perturbations of Schwarzschild like solutions in static coordinates (\ref{eq:line_element_static}) for any value of cosmological constant (here we reproduce well known linear results and then extend them to nonlinear orders), and perturbations of de Sitter spacetime in comoving coordinates (\ref{eq:line_element_comoving}) (giving linear scalar wave equation missing in \cite{Viaggiu} and then extending it to nonlinear orders). We conclude in Sec.~\ref{Sec:summary}. 
%%%%%%%%%%%%%%%%%%%%%%%%%%%%%%%%%%%%%%%%%%%%%%%%%%%%%%%%%%%%%%%%%%%%%%%%%%%%%%
%%%%%%%%%%%%%%%%%%%%%%%%%%%%%%%%%%%%%%%%%%%%%%%%%%%%%%%%%%%%%%%%%%%%%%%%%%%%%%
%%%%%%%%%%%%%%%%%%%%%%%%%%%%%%%%%%%%%%%%%%%%%%%%%%%%%%%%%%%%%%%%%%%%%%%%%%%%%%

%%%%%%%%%%%%%%%%%%%%%%%%%%%%%%%%%%%%%%%%%%%%%%%%%%%%%%%%%%%%%%%%%%%%%%%%%%%%%%
%%%%%%%%%%%%%%%%%%%%%%%%%%%%%%%%%%%%%%%%%%%%%%%%%%%%%%%%%%%%%%%%%%%%%%%%%%%%%%
%%%%%%%%%%%%%%%%%%%%%%%%%%%%%%%%%%%%%%%%%%%%%%%%%%%%%%%%%%%%%%%%%%%%%%%%%%%%%%
\section{General setup for gravitational perturbations in vacuum and Regge-Wheeler expansion into spherical harmonics}
\label{Sec:general}
%%%%%%%%%%%%%%%%%%%%%%%%%%%%%%%%%%%%%%%%%%%%%%%%%%%%%%%%%%%%%%%%%%%%%%%%%%%%%%
We have presented the general approach to gravitational perturbations in Sec.2 of \cite{r2017}; here we reproduce the most relevant formulas to make this work self contained. 

We are interested in solutions of vacuum Einstein equations 
\begin{equation}
\label{Eeqs}
R_{\mu \nu} -\kappa \frac{d}{\lsc^2} g_{\mu \nu} = 0,
\end{equation} 
where we allow for the nonzero value of the cosmological constant $\Lambda= \kappa \frac{d(d-1)}{2 \lsc^2}$; here $d$ is the number of spatial dimensions, and $\kappa=0, +1, -1$, for asymptotically Minkowski, de Sitter (dS) and anti-de Sitter (AdS) solutions respectively. Let the "bar" quantities stand for the exact solution of vacuum Einstein equations that we wish to perturb (i.e. Schwarzschild-- Minkowski, dS or AdS). Now let $g_{\mu\nu} = \bar g_{\mu \nu} + \delta g_{\mu\nu}$ or in matrix notation $g=\bar g + \delta g$, where $\delta g_{\mu\nu}$ will be expanded as
\vskip 0.1mm
\begin{equation}
\label{pertSer}
\delta g_{\mu\nu} = \sum_{1 \leq i} \pert{i}{h}_{\mu \nu} \, \epsilon^i 
\end{equation}
%\nopagebreak 
later on. Then we have:
%\enlargethispage{10mm}
%\pagebreak
\begin{widetext}
\begin{eqnarray}
\label{eq:gInverse}
g^{\alpha \beta} &=& \left(\bar g^{-1} - \bar g^{-1} \delta g \bar g^{-1} + \bar g ^{-1} \delta g \bar g ^{-1} \delta g \bar g ^{-1} - \dots \right)^{\alpha \beta} 
\nonumber\\
&=& \bar g^{\alpha \beta} + \delta g^{\alpha \beta} \,,
\\
\label{eq:GammaPert}
\Gamma^{\alpha}_{\mu \nu} &=& \bar \Gamma^{\alpha}_{\mu \nu} + \frac {1} {2} \left(\bar g^{-1} - \bar g^{-1} \delta g \bar g^{-1} + \bar g ^{-1} \delta g \bar g ^{-1} \delta g \bar g ^{-1} - \dots \right)^{\alpha \lambda} (\bar \nabla_{\mu} \delta g_{\lambda \nu} + \bar \nabla_{\nu} \delta g_{\lambda \mu} - \bar \nabla_{\lambda} \delta g_{\mu \nu}) 
\nonumber\\
&=& \bar \Gamma^{\alpha}_{\mu \nu} + \delta \Gamma^{\alpha}_{\mu \nu} \, ,
\\
\label{eq:RPert}
R_{\mu \nu} &=& \bar R_{\mu \nu} + \bar \nabla_{\alpha} \delta \Gamma^{\alpha}_{\mu \nu} - \bar \nabla_{\nu} \delta \Gamma^{\alpha}_{\alpha \mu} + \delta \Gamma^{\alpha}_{\alpha \lambda} \delta \Gamma^{\lambda}_{\mu \nu} - \delta \Gamma^{\lambda}_{\mu \alpha} \delta \Gamma^{\alpha}_{\lambda \nu} 
\nonumber\\
&=&  \bar R_{\mu \nu} +  \delta R_{\mu \nu} \, .
\end{eqnarray} 
\end{widetext}
Equation (\ref{eq:gInverse}) is straightforwardly obtained by recursive application of the formula $g^{\alpha \beta} = \bar g^{\alpha \beta} - \bar g^{\alpha \mu} \delta g_{\mu \nu} g^{\nu \beta}$ and then Eqs. (\ref{eq:GammaPert}) and (\ref{eq:RPert}) easily follow. Thus the Einstein equations read ("bar" quantities are solutions to Einstein equations)
\begin{equation}
\label{EeqsPert}
\delta R_{\mu \nu} - \kappa \frac{d}{\lsc^2} \delta g_{\mu \nu} = 0.
\end{equation} 
Plugging the expansion (\ref{pertSer}) into (\ref{EeqsPert}) and collecting terms at the same powers of $\epsilon$, we get the following hierarchy of equations:
\begin{equation}
\label{eq:pertEq}
\pert{i}{E}_{\mu\nu} := \Delta_L \pert{i}{h}_{\mu \nu} - \pert{i}{S}_{\mu \nu} = 0\, ,
\end{equation}
where the \textit{Lorentzian} Lichnerowicz operator $\Delta_L$ and the source terms read
\begin{widetext}
\begin{equation}
\Delta_L h_{\mu \nu} = \frac{1}{2} \left(- \bar \nabla^{\alpha} \bar \nabla_{\alpha} h_{\mu \nu} - \bar \nabla_{\mu} \bar \nabla_{\nu} h - 2 \bar R_{\mu \alpha \nu \beta} h^{\alpha \beta} + \bar \nabla_{\mu} \bar \nabla^{\alpha} h_{\nu \alpha} + \bar \nabla_{\nu} \bar \nabla^{\alpha} h_{\mu \alpha} \right) \,,  
\label{eq:Delta_L}
\end{equation}
\begin{align}
\pert{i}{S}_{\mu \nu} = \left[\epsilon^i \right] & \left\{ 
- (1/2) \bar \nabla_{\alpha} \left[ \left(- \bar g^{-1} \delta g \bar g^{-1} + \bar g ^{-1} \delta g \bar g ^{-1} \delta g \bar g ^{-1} - \dots \right)^{\alpha \lambda} \left(\bar \nabla_{\mu} \delta g_{\lambda \nu} + \bar \nabla_{\nu} \delta g_{\lambda \mu} - \bar \nabla_{\lambda} \delta g_{\mu \nu}\right)\right] \right.
\nonumber\\
& \hskip 1.4mm + (1/2) \bar \nabla_{\nu} \left[\left(- \bar g^{-1} \delta g \bar g^{-1} + \bar g ^{-1} \delta g \bar g ^{-1} \delta g \bar g ^{-1} - \dots \right)^{\alpha \lambda} \left(\bar \nabla_{\mu} \delta g_{\lambda \alpha} + \bar \nabla_{\alpha} \delta g_{\lambda \mu} - \bar \nabla_{\lambda} \delta g_{\mu \alpha}\right)\right]
\nonumber\\
& \left. \hskip 1.4mm - \delta \Gamma^{\alpha}_{\alpha \lambda} \delta \Gamma^{\lambda}_{\mu \nu} + \delta \Gamma^{\lambda}_{\mu \alpha} \delta \Gamma^{\alpha}_{\lambda \nu} \right\} \,,
\label{eq:source_definition}
\end{align} 
\end{widetext}
and $\left[ \epsilon^i \right]f$ denotes the coefficient at
$\epsilon^i$ in the (formal) power series expansion of $f=\sum_i f_i
\epsilon^i$. The allowed nonzero value of cosmological constant does
not enter the formulas (\ref{eq:pertEq})-(\ref{eq:source_definition})
explicitly, but is implicitly present in the "bar" quantities. From
now on we restrict ourselves to $d=3$ spatial dimensions and use RW expansion
into scalar, vector and tensor spherical harmonics \cite{RW} (see also
Sec.2 in \cite{Nollert} and appendix B in \cite{Mukohyama}). In
particular, in $3+1$ dimensions any tensor $T_{\mu\nu}$ can be split
into seven polar and three axial components. For any tensor
$T_{\mu\nu}$ its polar components expanded into (one scalar-, one
vector- and two tensor-type) polar spherical harmonics at axial symmetry
read \footnote{Here and in the following scalar, vector and tensor spherical harmonics refer merely to functions into which scalar, vector and tensor quantities on a sphere can be expanded, i.e. in the sense used in \cite{Mukohyama}, and not in the sense of classification given in \cite{KI}.}
\begin{widetext}
\begin{align}
\label{eq:polarT_ab}
T_{ab}(t,r,\theta) = & \sum_{0 \leq \ell} T_{\ell\,\,ab} (t,r) P_{\ell}(\cos\theta), \quad a,b=0,1 \, , 
\\
\label{eq:polarT_a2}
T_{a2}(t,r,\theta) = & \sum_{1 \leq \ell} T_{\ell\,\,a2} (t,r) \partial_\theta P_{\ell}(\cos\theta), \quad a=0,1 \, , 
\\
\label{eq:polarT+}
\frac{1}{2}\left(T_{22}(t,r,\theta) + \frac{T_{33}(t,r,\theta)}{\sin^{2}\!\theta}\right) = & \sum_{0 \leq \ell} T_{\ell\,\,+} (t,r) P_{\ell}(\cos\theta) \, ,
\\
%\label{eq:S-}
\label{eq:polarT-}
\frac{1}{2}\left(T_{22}(t,r,\theta) - \frac{T_{33}(t,r,\theta)}{\sin^{2}\!\theta}\right) = & \sum_{2 \leq \ell} T_{\ell\,\,-} (t,r) \left(-\ell(\ell+1) P_{\ell}(\cos\theta) - 2 \cot{\theta} \partial_{\theta}P_{\ell}(\cos\theta)\right) \, ,
\end{align}
\end{widetext}
where $P_{\ell}$ are Legendre polynomials. For any tensor $T_{\mu\nu}$
its axial components expanded into (one vector- and one tensor-type) axial
spherical harmonics at axial symmetry read
\begin{widetext}
\begin{align}
\label{eq:axialT_a3}
T_{a3}(t,r,\theta) = & \sum_{1 \leq \ell} T_{\ell\,\,a3} (t,r) \sin \theta \partial_{\theta} P_{\ell}(\cos\theta), \quad a=0,1 \, , 
\\
\label{eq:axialT_23}
T_{23}(t,r,\theta) = & \sum_{2 \leq \ell} T_{\ell\,\,23} (t,r) \left(- 2 \cos \theta \partial_{\theta} P_\ell(\cos\theta) - \ell(\ell+1) \sin \theta P_\ell(\cos\theta) \right) \, .
\end{align}
\end{widetext}
Accordingly, in $3+1$ dimensions any vector $V_{\mu}$ can be split
into three polar and one axial component. For any vector $V_{\mu}$ its
polar components expanded into (one scalar- and one vector-type) polar
spherical harmonics at axial symmetry read
\begin{align}
\label{eq:polarV_a}
V_{a}(t,r,\theta) = & \sum_{0 \leq \ell} V_{\ell\,\,a} (t,r) P_{\ell}(\cos\theta), \quad a = 0,1 \, , 
\\
\label{eq:polarV_2}
V_{2}(t,r,\theta) = & \sum_{1 \leq \ell} V_{\ell\,\,2} (t,r) \partial_\theta P_{\ell}(\cos\theta) \, . 
\end{align}
For any vector $V_{\mu}$ its axial component expanded into (one
vector-type) axial spherical harmonics at axial symmetry reads
\begin{align}
\label{eq:axialV_3}
V_{3}(t,r,\theta) = & \sum_{1 \leq \ell} V_{\ell\,\,3} (t,r) \sin \theta \partial_{\theta} P_{\ell}(\cos\theta) \, . 
\end{align}
In what follows, the symbols $\pert{i}{h}_{\ell\,\,\mu\nu}$, $\pert{i}{S}_{\ell\,\,\mu\nu}$, $\pert{i}{E}_{\ell\,\,\mu\nu}$ and $\Delta_L \pert{i}{h}_{\ell\,\,\mu\nu}$ appear in expansion of \textit{tensors} $\pert{i}{h}_{\mu\nu}$, $\pert{i}{S}_{\mu\nu}$, $\pert{i}{E}_{\mu\nu}$ and $\Delta_L \pert{i}{h}_{\mu\nu}$ according to (\ref{eq:polarT_ab}-\ref{eq:axialT_23}) respectively (cf. (\ref{pertSer},\ref{eq:pertEq})). The symbols $\pert{i}{\zeta}_{\ell\,\,\mu}$ 
%and $\pert{i}{\eta}_{\ell\,\,\mu}$ 
appear in expansion of polar gauge \textit{vectors} $\pert{i}{\zeta}_{\mu}$ 
%and axial gauge \textit{vectors} $\pert{i}{\eta}_{\mu}$ 
according to (\ref{eq:polarV_a},\ref{eq:polarV_2}) 
%(\ref{eq:polarV_a}-\ref{eq:axialV_3}) respectively 
(see below for the usage of these gauge vectors). 

After separating angular dependence, the system of ten perturbative Einstein equations (\ref{eq:pertEq}) splits into the system of seven equations for polar (alternatively called scalar or even) type perturbations and the system of three equations for axial (alternatively called vector or odd) type perturbations \cite{RW, Nollert}. The polar and axial parts decouple at linear order but generally mix at higher orders. The axial symmetry is exceptional in this respect, as if we
excite only polar perturbations at the linear order we stay in the polar sector at all higher orders as well, while starting with only axial perturbations at linear order results in axial perturbations at all odd orders and polar perturbations at all even orders of perturbation expansion. The most general gauge vector (for given spherical harmonics index $\ell$) splits accordingly into polar and axial
parts parametrized by three and one function respectively. It turns out that the Lichnerowicz operator (\ref{eq:Delta_L}) depends only on four/two RW gauge invariant variables in polar/axial sector, while the sources (\ref{eq:source_definition}) contain also gauge degrees of freedom.
%%%%%%%%%%%%%%%%%%%%%%%%%%%%%%%%%%%%%%%%%%%%%%%%%%%%%%%%%%%%%%%%%%%%%%%%%%%%%%
%%%%%%%%%%%%%%%%%%%%%%%%%%%%%%%%%%%%%%%%%%%%%%%%%%%%%%%%%%%%%%%%%%%%%%%%%%%%%%
%%%%%%%%%%%%%%%%%%%%%%%%%%%%%%%%%%%%%%%%%%%%%%%%%%%%%%%%%%%%%%%%%%%%%%%%%%%%%%

%%%%%%%%%%%%%%%%%%%%%%%%%%%%%%%%%%%%%%%%%%%%%%%%%%%%%%%%%%%%%%%%%%%%%%%%%%%%%%
%%%%%%%%%%%%%%%%%%%%%%%%%%%%%%%%%%%%%%%%%%%%%%%%%%%%%%%%%%%%%%%%%%%%%%%%%%%%%%
%%%%%%%%%%%%%%%%%%%%%%%%%%%%%%%%%%%%%%%%%%%%%%%%%%%%%%%%%%%%%%%%%%%%%%%%%%%%%%
\section{Perturbations of Schwarzschild like solutions in static coordinates}
\label{Sec:Schwarzschild}
%%%%%%%%%%%%%%%%%%%%%%%%%%%%%%%%%%%%%%%%%%%%%%%%%%%%%%%%%%%%%%%%%%%%%%%%%%%%%%
In this section we generalize the results of the paper \cite{r2017} to the perturbations of spherically symmetric space-times in static, Schwarzschild like coordinates:
\begin{equation}
\label{eq:line_element_static}
ds^2 = -\fsc dt^2 + \fsc^{-1} dr^2 + r^2 d\Omega^2_2 \, ,
\end{equation}
with 
\begin{equation}
\label{eq:fsc_static}
\fsc \equiv \fsc(r) = \left(1 -\kappa \, r^2/\lsc^2 - 2M /
r\right) \, .  
\end{equation}
This line element is a solution to vacuum Einstein equations (\ref{Eeqs}) and $r^2 \fsc'' = 2\fsc -2$ holds. We focus on polar perturbations at axial symmetry as the most illustrative example and we show how general ideas of Sec.\ref{Sec:Intro} work in practice for gravitational perturbations of zero order solution (\ref{eq:line_element_static}). In axial symmetry for polar perturbations we have
\begin{equation}
  \label{eq:hPolar}
  \left( \pert{i}{h}_{\alpha\beta} \right) = 
\left(\begin{array}{cccc}
    \pert{i}{h}_{00} & \pert{i}{h}_{01} & \pert{i}{h}_{02} & 0 \\
    \pert{i}{h}_{01} & \pert{i}{h}_{11} & \pert{i}{h}_{12} & 0 \\
    \pert{i}{h}_{02} & \pert{i}{h}_{12} & \pert{i}{h}_{22} & 0 \\
    0 & 0 & 0 & \pert{i}{h}_{33}  
\end{array}\right) \, ,
\end{equation}
with (cf. (\ref{eq:polarT_ab}-\ref{eq:polarT-}))
\begin{align}
\label{eq:h00}
\pert{i}{h}_{\ell\,\,00} = & \pert{i}{f}_{\ell\,\,00} + 2 \partial_t \pert{i}{\zeta}_{\ell\,\,0} - \fsc \, \fsc' \, \pert{i}{\zeta}_{\ell\,\,1} \, ,
\\
\pert{i}{h}_{\ell\,\,11} = & \pert{i}{f}_{\ell\,\,11} + 2 \partial_r \pert{i}{\zeta}_{\ell\,\,1} + \frac{\fsc'}{\fsc} \pert{i}{\zeta}_{\ell\,\,1} \, ,
\\
\pert{i}{h}_{\ell\,\,01} = & \pert{i}{f}_{\ell\,\,01} + \partial_r \pert{i}{\zeta}_{\ell\,\,0} + \partial_t \pert{i}{\zeta}_{\ell\,\,1} - \frac{\fsc'}{\fsc} \pert{i}{\zeta}_{\ell\,\,0} \, ,
\\
\pert{i}{h}_{\ell\,\,02} = & \pert{i}{\zeta}_{\ell\,\,0} + \partial_t \pert{i}{\zeta}_{\ell\,\,2} \, ,
\\
\pert{i}{h}_{\ell\,\,12} = & \pert{i}{\zeta}_{\ell\,\,1} - \frac{2}{r}\pert{i}{\zeta}_{\ell\,\,2} + \partial_r \pert{i}{\zeta}_{\ell\,\,2} \, ,
\\
%\end{align}
%\begin{align}
\pert{i}{h}_{\ell\,\,+}  = & r^2 \pert{i}{f}_{\ell\,\,+} + 2 r \fsc \, \pert{i}{\zeta}_{\ell\,\,1} - \ell(\ell+1) \pert{i}{\zeta}_{\ell\,\,2} \, ,
\\
\label{eq:h-}
\pert{i}{h}_{\ell\,\,-} = & \pert{i}{\zeta}_{\ell\,\,2} \, , 
\end{align}
where $\pert{i}{\zeta}_{\ell\,\,0}$, $\pert{i}{\zeta}_{\ell\,\,1}$, $\pert{i}{\zeta}_{\ell\,\,2}$ polar components define the $i$-th order polar gauge vector $\pert{i}{\zeta}_{\mu}$ (cf. (\ref{eq:polarV_a},\ref{eq:polarV_2})) and $\pert{i}{f}_{\ell\,\,00}(t,r)$, $\pert{i}{f}_{\ell\,\,11}(t,r)$, $\pert{i}{f}_{\ell\,\,01}(t,r)$, $\pert{i}{f}_{\ell\,\,+}(t,r)$ are Regge-Wheeler variables \cite{RW, Nollert}, being gauge invariant with respect to gauge transformations induced by $\pert{j}{\zeta}_{\mu}$ with $j \geq i$, i.e. gauge transformations of the form
\begin{equation}
\label{eq:gaugeTransform}
\sum_{1 \leq i} \pert{i}{h}_{\mu \nu} \, \epsilon^i \rightarrow \sum_{1 \leq i} \pert{i}{h}_{\mu \nu} \, \epsilon^i + \epsilon^j\mathcal{L}_{\pert{j}{\zeta}} \bar g_{\mu \nu} + \mathcal{O}\left(\epsilon^{j+1}\right).
\end{equation}
Of course, $\pert{i}{f}_{\ell\,\,00}$, $\pert{i}{f}_{\ell\,\,11}$, $\pert{i}{f}_{\ell\,\,01}$, $\pert{i}{f}_{\ell\,\,+}$ so defined are not gauge invariant in general; they change under gauge transformations induced by ${}^{(j)}\zeta^{\mu}$ with $j < i$ (cf. Bruni \textit{et al.} \cite{BMMS}). The RW gauge corresponds to setting $\pert{i}{\zeta}_{\ell\,\,0} = \pert{i}{\zeta}_{\ell\,\,1} = \pert{i}{\zeta}_{\ell\,\,2} = 0$ in (\ref{eq:h00}-\ref{eq:h-}). At each nonlinear order ($i>1$) sources have the form
\begin{widetext}
\begin{equation}
  \label{eq:S_order_i}
  \left( \pert{i}{S}_{\alpha\beta} \right) =
\left(\begin{array}{cccc}
    \pert{i}{S}_{00} & \pert{i}{S}_{01} & \pert{i}{S}_{02} & 0 \\
    \pert{i}{S}_{01} & \pert{i}{S}_{11} & \pert{i}{S}_{12} & 0 \\
    \pert{i}{S}_{02} & \pert{i}{S}_{12} & \pert{i}{S}_{22} & 0 \\
    0 & 0 & 0 & \pert{i}{S}_{33} \\
\end{array}\right) \, ,
\end{equation}
with the components expanded according to (\ref{eq:polarT_ab}-\ref{eq:polarT-}). As already indicated in Sec.\ref{Sec:Intro}, it is crucial to note that the sources in the polar sector of perturbative Einstein equations are not independent but fulfill three identities:
\begin{align}
\label{eq:zero0}
\pert{i}{\Nsc}_{\ell\,\,0} & := \frac{1}{2} \left( \frac{1}{\fsc} \partial_t \pert{i}{S}_{\ell\,\,00} + \fsc \partial_t \pert{i}{S}_{\ell\,\,11}\right) + \frac{1}{r^2}\partial_t \pert{i}{S}_{\ell\,\,+} - \fsc \partial_r \pert{i}{S}_{\ell\,\,01} - \frac{2 \fsc + r \, \fsc'}{r} \pert{i}{S}_{\ell\,\,01} + \frac{\ell(\ell+1)}{r^2} \pert{i}{S}_{\ell\,\,02} = 0\, ,
\\
\label{eq:zero1}
\pert{i}{\Nsc}_{\ell\,\,1} & := \frac{1}{2} \left(\frac{1}{\fsc} \partial_r \pert{i}{S}_{\ell\,\,00} + \fsc \partial_r \pert{i}{S}_{\ell\,\,11} \right) - \frac{1}{r^2}\partial_r \pert{i}{S}_{\ell\,\,+} - \frac{1}{\fsc} \partial_t \pert{i}{S}_{\ell\,\,01} + \frac{2 \fsc + r \, \fsc'}{r} \pert{i}{S}_{\ell\,\,11} - \frac{\ell(\ell+1)}{r^2} \pert{i}{S}_{\ell\,\,12} = 0\, ,
\\
\label{eq:zero2}
\pert{i}{\Nsc}_{\ell\,\,2} & := \frac{1}{2} \left( \frac{1}{\fsc} \, \pert{i}{S}_{\ell\,\,00} - \fsc \, \pert{i}{S}_{\ell\,\,11}\right) - \frac{1}{\fsc} \partial_t \pert{i}{S}_{\ell\,\,02} + \fsc \partial_r \pert{i}{S}_{\ell\,\,12} + \frac{2\fsc + r \, \fsc'}{r} \pert{i}{S}_{\ell\,\,12} - \frac{(\ell-1)(\ell+2)}{r^2} \pert{i}{S}_{\ell\,\,-} = 0\, .
\end{align}
\end{widetext}
These identities are easily easily obtained by taking the background divergence of (\ref{eq:pertEq}): $\pert{i}{\Nsc}_{\ell\,\,\nu} = 0$ (with $\nu=0,1,2$) follows from (three) polar components of $\bar \nabla^{\mu} \pert{i}{E}_{\mu\nu}=0$.

The polar components of the system of perturbative Einstein equations (\ref{eq:pertEq}) are expanded according to (\ref{eq:polarT_ab}-\ref{eq:polarT-}). The gauge degrees of freedom enter $\pert{i}{E}_{\ell\,\,\mu\nu}$ only through the source terms $\pert{i}{S}_{\ell\,\,\mu\nu}$ ($\pert{i}{S}_{\ell\,\,\mu\nu}$ depend on gauge functions ${}^{(j)}\zeta_{\ell\,\,0}$, ${}^{(j)}\zeta_{\ell\,\,1}$, ${}^{(j)}\zeta_{\ell\,\,2}$ with $j<i$, that is on the gauge choices made in previous steps). On the other hand, $\Delta_L \pert{i}{h}_{\ell\,\,\mu\nu}$ components are given in terms of Regge-Wheeler gauge invariant variables $\pert{i}{f}_{\ell\,\,00}$, $\pert{i}{f}_{\ell\,\,11}$, $\pert{i}{f}_{\ell\,\,01}$, $\pert{i}{f}_{\ell\,\,+}$ only. Thus, we solve (\ref{eq:pertEq}) at order $i$ for Regge-Wheeler variables $\pert{i}{f}_{\ell\,\,00}$, $\pert{i}{f}_{\ell\,\,11}$, $\pert{i}{f}_{\ell\,\,01}$, $\pert{i}{f}_{\ell\,\,+}$ and then recover the required asymptotic behavior of metric perturbations (\ref{eq:hPolar}) with a suitable gauge transformation. We note that
\begin{equation}
\label{eq:algebraicE-}
\pert{i}{E}_{\ell\,\,-} = \frac{1}{4} \left( \frac{1}{\fsc} \, \pert{i}{f}_{\ell\,\,00}  - \fsc \, \pert{i}{f}_{\ell\,\,11} \right) - \pert{i}{S}_{\ell\,\,-} = 0
\end{equation}
sets the purely algebraic relation between $\pert{i}{f}_{\ell\,\,00}$ and $\pert{i}{f}_{\ell\,\,11}$, while the combination
\begin{widetext}
\begin{align}
0 =& - \frac{1}{2} \left( \frac{1}{\fsc} \, \pert{i}{E}_{\ell\,\,00} - \fsc \, \pert{i}{E}_{\ell\,\,11} \right) +  \frac{1}{\fsc} \partial_t \pert{i}{E}_{\ell\,\,02}  - \fsc \partial_r \pert{i}{E}_{\ell\,\,12} - \frac{2 \fsc + r \, \fsc'}{r} \pert{i}{E}_{\ell\,\,12} + \frac{(\ell-1)(\ell+2)}{r^2} \pert{i}{E}_{\ell\,\,-}  
\nonumber\\
\label{eq:sourceIdentity}
& \equiv \frac{1}{2} \left( \frac{1}{\fsc} \, \pert{i}{S}_{\ell\,\,00} - \fsc \, \pert{i}{S}_{\ell\,\,11}^{(i)} \right) - \frac{1}{\fsc} \partial_t \pert{i}{S}_{\ell\,\,02}  + \fsc \partial_r \pert{i}{S}_{\ell\,\,12} + \frac{2\fsc + r \, \fsc' }{r} \pert{i}{S}_{\ell\,\,12} - \frac{(\ell-1)(\ell+2)}{r^2} \pert{i}{S}_{\ell\,\,-}
\end{align}
\end{widetext}
reduces to the identity (\ref{eq:zero2}) fulfilled by the sources. Thus after eliminating $\pert{i}{f}_{\ell\,\,00}$ from (\ref{eq:algebraicE-}) we are left with three RW gauge invariant variables $\pert{i}{f}_{\ell\,\,11}$, $\pert{i}{f}_{\ell\,\,01}$, $\pert{i}{f}_{\ell\,\,+}$ to satisfy five linearly independent equations: $\pert{i}{E}_{\ell\,\,00} = \pert{i}{E}_{\ell\,\,01} = \pert{i}{E}_{\ell\,\,+} = \pert{i}{E}_{\ell\,\,02} = \pert{i}{E}_{\ell\,\,12} = 0$. 

Next we show that we can easily fulfill all perturbative Einstein equations (\ref{eq:pertEq}) introducing master scalar variables along the lines of Sec.\ref{Sec:Intro}. First, we assume that these master scalar variables $\pert{i}{\Phi}_{\ell}^{\Psc} \equiv \pert{i}{\Phi}_{\ell}^{\Psc} (t,r)$ fulfill inhomogeneous (at nonlinear orders), linear scalar wave equation on the zero-order solution (\ref{eq:line_element_static}), with a potential:
\begin{equation}
\label{eq:scalar_wave}
\tilde \Box_{\ell}^{\Psc} \pert{i}{\Phi}_{\ell}^{\Psc} := r \left(-\bar \Box + V_{\ell}^{\Psc} \right) \frac{\pert{i}{\Phi}_{\ell}^{\Psc}}{r} = \pert{i}{\tilde S}_{\ell}^{\Psc}\,,
\end{equation}
where $\bar \Box$ is the scalar wave operator for the metric defined by the line element (\ref{eq:line_element_static}), and the potentials $V_{\ell}^{\Psc}$ and inhomogenities (scalar sources) $\pert{i}{\tilde S}_{\ell}^{\Psc}$ will be defined below (at linear order $\pert{1}{\tilde S}_{\ell}^{\Psc} \equiv 0$ and the wave eq. (\ref{eq:scalar_wave}) is homogeneous). It is easy to check that for the line element (\ref{eq:line_element_static})
\begin{equation}
\label{eq:scalar_wave_static}
\tilde \Box_{\ell}^{\Psc} \pert{i}{\Phi}_{\ell}^{\Psc} = \left(\frac{\partial_{tt}}{\fsc} - \fsc \, \partial_{rr} - \fsc' \, \partial_{r} + \frac{\fsc'}{r} + V_{\ell}^{\Psc} \right) \pert{i}{\Phi}_{\ell}^{\Psc}\, .
\end{equation}
Then we express RW variables $\pert{i}{f}_{\ell\,\,11}$, $\pert{i}{f}_{\ell\,\,01}$ and $\pert{i}{f}_{\ell\,\,+}$ as linear combinations of master scalar variables $\pert{i}{\Phi}_{\ell}^{\Psc}$ and their (partial) derivatives up to the second order, and source functions to be defined below. The potential $V_{\ell}^{\Psc}$ in (\ref{eq:scalar_wave}) and linear combination of master scalar variables and their partial derivatives are chosen to satisfy the homogeneous part of perturbative Einstein equations (\ref{eq:pertEq}), while the source functions are chosen to satisfy inhomogenities of these equations. Indeed, there is a unique choice for the potential $V_{\ell}^{\Psc}$ in (\ref{eq:scalar_wave}), and a unique linear combination of master scalar variables (up to a multiplicative factor) that solves homogeneous part of (\ref{eq:pertEq}) identically, namely $\Delta_L \pert{i}{h}_{\mu \nu} \equiv 0$ if and only if
\begin{widetext}
\begin{equation}
\label{eq:potential_static_polar}
V_{\ell}^{\Psc} = \frac{\ell(\ell+1)}{r^2} - \frac{\fsc'}{r} + \underbrace{\left( 2 \fsc - r \fsc' - 2 \right)}_{-6M/r} \frac{2 \fsc \left(r \fsc' - 2 \right) - \left(r \fsc' \right)^2 + \ell^2(\ell+1)^2}{r^2 \left( 2 \fsc - r \fsc' - \ell(\ell+1) \right)^2}
\end{equation}
\begin{align}
\label{eq:f+}
\pert{i}{f}_{\ell\,\,+} = & \fsc \partial_{r} \pert{i}{\Phi}^{\Psc}_{\ell} + \frac{1}{r}\left( \frac{\ell(\ell+1)}{2} - \frac{2 \fsc - r \fsc' - 2}{2 \fsc - r \fsc' - \ell(\ell+1)} \fsc \right) \pert{i}{\Phi}_{\ell}^{\Psc} + \pert{i}{\alpha}_{\ell}\, ,
\\
\label{eq:f11}
\pert{i}{f}_{\ell\,\,11} = & r \partial_{rr} \pert{i}{\Phi}_{\ell}^{\Psc} + \left( 2 - \frac{1}{\fsc} - \left( 2 \fsc - r \fsc' - 2 \right) \frac{4 \fsc - r \fsc' - \ell(\ell+1)}{2 \fsc \left( 2 \fsc - r \fsc' - \ell(\ell+1) \right)} \right) \partial_{r} \pert{i}{\Phi}_{\ell}^{\Psc} - \frac{r}{2 \fsc} \left( V_{\ell} + \frac{\fsc'}{r} \right) \pert{i}{\Phi}_{\ell}^{\Psc} + \pert{i}{\beta}_{\ell}  
\\
\label{eq:f01}
\pert{i}{f}_{\ell\,\,01} = & r \partial_{tr} \pert{1}{\Phi}^{\Psc}_{\ell} + \left( \frac{1}{\fsc} - \left( 2 \fsc - r \fsc' - 2 \right) \frac{r \fsc' + \ell(\ell+1)}{2 \fsc \left( 2 \fsc - r \fsc' - \ell(\ell+1) \right)} \right) \partial_{t} \pert{i}{\Phi}_{\ell}^{\Psc}  + \pert{i}{\gamma}_{\ell}\, ,
\end{align}
\end{widetext}
and we set the source functions in (\ref{eq:f+}-\ref{eq:f01}) and the scalar source in (\ref{eq:scalar_wave}) to zero: $\pert{i}{\alpha}_{\ell} = \pert{i}{\beta}_{\ell} = \pert{i}{\gamma}_{\ell} = 0 = \pert{i}{\tilde S}_{\ell}^{\Psc}$ (their nonzero values will take care for the inhomogenities in perturbation Einstein equations (\ref{eq:pertEq})). We used here the scalar wave equation (\ref{eq:scalar_wave}) to eliminate from our expressions all partial derivatives of $\pert{i}{\Phi}_{\ell}^{\Psc}$ with respect to time, higher then the first order, and $\pert{i}{f}_{\ell\,\,00}$ is fixed from (\ref{eq:algebraicE-})). In other words, eqs. (\ref{eq:algebraicE-},\ref{eq:scalar_wave},\ref{eq:potential_static_polar}-\ref{eq:f01}) constitute the solution of perturbative Einstein equations (\ref{eq:pertEq}) at the linear order $(i=1)$ and (\ref{eq:potential_static_polar}) encompasses the celebrated Zerilli potential \cite{Zerilli}, and its analogues for Schwarzschild-dS \cite{MM} and Schwarzschild-AdS \cite{CL} black holes. Now, the scalar sources $\pert{i}{\tilde S}_{\ell}^{\Psc}$ and source functions $\pert{i}{\alpha}_{\ell}$, $\pert{i}{\beta}_{\ell}$, $\pert{i}{\gamma}_{\ell}$ can be set to fulfill perturbative Einstein equations at any nonlinear order. To find $\pert{i}{\tilde S}_{\ell}^{\Psc}$, the relations (\ref{eq:f+}-\ref{eq:f01}) have to be inverted for $\pert{i}{\Phi}_{\ell}^{\Psc}$ itself. Moreover, such inversion is necessary to provide initial data for the scalar wave equation (\ref{eq:scalar_wave}), once the metric perturbations are prescribed at some initial time slice $t=\,$const. In general such inversion would not be uniquely defined, however it becomes uniquely defined if we wish to provide the initial data for (\ref{eq:scalar_wave}), i.e. $\pert{i}{\Phi}_{\ell}^{\Psc}$ and $\partial_t \pert{i}{\Phi}_{\ell}^{\Psc}$ at some initial time-slice in terms of the first and the second fundamental forms of an initial data hypersurface. To achieve it, $\pert{i}{\Phi}_{\ell}^{\Psc}$ should be given in terms of linear combination of $\pert{i}{f}_{\ell\,\,11}$, $\pert{i}{f}_{\ell\,\,+}$ and their radial derivatives only. There is a uniquely defined such linear combination:
\begin{equation}
\label{eq:Phi_def}
\pert{i}{\Phi}^{\Psc}_{\ell} = \frac{2 r}{\ell(\ell+1)} \left( \pert{i}{f}_{\ell\,\,+}  + 2 \fsc \, \frac{\fsc \, \pert{i}{f}_{\ell\,\,11} - r \partial_r \pert{i}{f}_{\ell\,\,+}}{\ell(\ell+1) - 2 \fsc + r \fsc'}\right) \,.
\end{equation}
Indeed, at linear order, upon substituting (\ref{eq:f+},\ref{eq:f11}), with (\ref{eq:scalar_wave}) fulfilled, (\ref{eq:Phi_def}) becomes an identity. The definition (\ref{eq:Phi_def}) coincides in the Schwarzschild case with the Moncrief's definition of master scalar variables for polar perturbations \cite{Moncrief}. We take (\ref{eq:Phi_def}) as a \textit{definition} of master scalar variables for polar perturbations at any (nonlinear) order. This definition
determines the scalar source $\pert{i}{\tilde S}_{\ell}^{\Psc}$ in the inhomogeneous scalar wave equation (\ref{eq:scalar_wave}), namely substituting (\ref{eq:Phi_def}) in (\ref{eq:scalar_wave_static}) we get
\begin{widetext}
\begin{align}
& \tilde \Box_{\ell} \pert{i}{\Phi}^{\Psc}_{\ell} = \frac {4 r^2} {(\ell-1)\ell(\ell+1)(\ell+2)} 
\nonumber\\
\times & \left( \frac{\fsc}{r} \left( \fsc \Delta_L \pert{i}{h}_{\ell\,\,11} - \frac{1}{\fsc} \Delta_L \pert{i}{h}_{\ell\,\,00} \right) + \frac{(\ell-1)(\ell+2) - 2 \left(3\fsc - 2\right)}{r^3} \Delta_L \pert{i}{h}_{\ell\,\,+}  - 2 \fsc \partial_r \left( \Delta_L \pert{i}{h}_{\ell\,\,+} / r^2 \right) \right.
\nonumber\\
& \hskip 1mm - \frac{2 \ell (\ell+1)}{r^2} \fsc \Delta_L \pert{i}{h}_{\ell\,\,12} + \frac{(\ell-1)\ell(\ell+1)(\ell+2)}{r^3} \Delta_L \pert{i}{h}_{\ell\,\,-}  - \frac{2 \fsc - r \fsc' - 2}{2 \fsc - r \fsc' - \ell(\ell+1)}
\nonumber\\
& \times \left( \frac{\fsc}{r} \left( \fsc \Delta_L \pert{i}{h}_{\ell\,\,11} - \frac{1}{\fsc} \Delta_L \pert{i}{h}_{\ell\,\,00} \right) - \frac{\fsc (\ell-1)(\ell+2)}{r\left( 2 \fsc - r \fsc' - \ell(\ell+1) \right)} \left( \fsc \Delta_L \pert{i}{h}_{\ell\,\,11} + \frac{1}{\fsc} \Delta_L \pert{i}{h}_{\ell\,\,00} \right) \right.
\nonumber\\
& - 2 \frac{3 \fsc \left( 2 \fsc - r \fsc' - 2 \right) - \ell(\ell+1) \left( 2 \fsc - r \fsc' - \ell(\ell+1) \right) - 2 (\ell-1)(\ell+1) \fsc}{r^3 \left( 2 \fsc - r \fsc' - \ell(\ell+1) \right)} \Delta_L \pert{i}{h}_{\ell\,\,+} 
\nonumber\\
& \left.\left. - 2 \fsc \partial_r \left( \Delta_L \pert{i}{h}_{\ell\,\,+} / r^2 \right) - \frac{2 \ell (\ell+1)}{r^2} \fsc \Delta_L \pert{i}{h}_{\ell\,\,12} \right) \right)\, .
\label{eq:proto_sw_source}
\end{align}  
Thus at higher orders ($i \geq 2$) the definition (\ref{eq:Phi_def}) leads to
\begin{align}
\label{eq:source_scalar_wave}
\pert{i}{\tilde S}^{\Psc}_{\ell} = & \frac {4 r^2} {(\ell-1)\ell(\ell+1)(\ell+2)} 
\nonumber\\
\times & \left( \frac{\fsc}{r} \left( \fsc \, \pert{i}{S}_{\ell\,\,11} - \frac{1}{\fsc} \, \pert{i}{S}_{\ell\,\,00} \right) + \frac{(\ell-1)(\ell+2) - 2 \left( 3 \fsc - 2 \right)}{r^3} \pert{i}{S}_{\ell\,\,+} - 2 \fsc \partial_r \left( \left. \pert{i}{S}_{\ell\,\,+} \right/ r^2 \right) \right.
\nonumber\\
& \hskip 1mm - \frac{2 \ell (\ell+1)}{r^2} \fsc \, \pert{i}{S}_{\ell\,\,12} + \frac{(\ell-1)\ell(\ell+1)(\ell+2)}{r^3} \pert{i}{S}_{\ell\,\,-} - \frac{2 \fsc - r \fsc' - 2}{2 \fsc - r \fsc' - \ell(\ell+1)}
\nonumber\\
& \times \left( \frac{\fsc}{r} \left( \fsc \pert{i}{S}_{\ell\,\,11} - \frac{1}{\fsc} \pert{i}{S}_{\ell\,\,00} \right) - \frac{\fsc (\ell-1)(\ell+2)}{r\left( 2 \fsc - r \fsc' - \ell(\ell+1) \right)} \left( \fsc \pert{i}{S}_{\ell\,\,11} + \frac{1}{\fsc} \pert{i}{S}_{\ell\,\,00} \right) \right.
\nonumber\\
& - 2 \frac{3 \fsc \left( 2 \fsc - r \fsc' - 2 \right) - \ell(\ell+1) \left( 2 \fsc - r \fsc' - \ell(\ell+1) \right) - 2 (\ell-1)(\ell+1) \fsc}{r^3 \left( 2 \fsc - r \fsc' - \ell(\ell+1) \right)} \pert{i}{S}_{\ell\,\,+} 
\nonumber\\
& \left.\left. - 2 \fsc \partial_r \left( \pert{i}{S}_{\ell\,\,+} / r^2 \right) - \frac{2 \ell (\ell+1)}{r^2} \fsc \pert{i}{S}_{\ell\,\,12} \right) \right)\, ,
\end{align}
\end{widetext}
for $\ell \geq 2$ (here the $\ell=0$ and $\ell=1$ cases have to be treated separately). We note that, even though the right hand side of eq. (\ref{eq:proto_sw_source}) is not uniquely defined in terms of linear combination of $\Delta_L \pert{i}{h}_{\ell\,\,\mu\nu}$ and their derivatives, the resulting formula (\ref{eq:source_scalar_wave}) is unique up to linear combination of the identities (\ref{eq:zero0}-\ref{eq:zero2}) and their derivatives. Finally, we are ready to set the source functions $\pert{i}{\alpha}_{\ell}$, $\pert{i}{\beta}_{\ell}$ and $\pert{i}{\gamma}_{\ell}$ in (\ref{eq:f+}-\ref{eq:f01}). They are set in such a way that all perturbative Einstein equations (\ref{eq:pertEq}) are satisfied and (\ref{eq:Phi_def}) becomes the identity upon substituting (\ref{eq:f+},\ref{eq:f11}) with (\ref{eq:scalar_wave}) fulfilled, also at nonlinear order. It is crucial to note, that without identities (\ref{eq:zero0}-\ref{eq:zero2}) it would be impossible to meet all these nonlinear requirements simultaneously. Moreover, it is worth to note that due to the identities (\ref{eq:zero0}-\ref{eq:zero2}), the solutions for $\pert{i}{\alpha}_{\ell}$, $\pert{i}{\beta}_{\ell}$ and $\pert{i}{\gamma}_{\ell}$ can be found in purely algebraic way, without solving any (partial) differential equations. Namely, we write down the source functions $\pert{i}{\alpha}_{\ell}$, $\pert{i}{\beta}_{\ell}$ and $\pert{i}{\gamma}_{\ell}$ as linear combination of the sources $\pert{i}{S}_{\ell\,\,\mu\nu}$ and their first derivatives. Fixing $3 \times 7 \times 3 = 63$ function coefficients of these linear combinations is a technical task with \textit{Mathematica}. It turns out that, out of these $63$ functions, $54$ functions are uniquely fixed in terms of $9$ free functions, and moreover, in the resulting expressions, the coefficients of $9$ free functions are identically zero, due to the identities (\ref{eq:zero0}-\ref{eq:zero2}). Thus we can put them to zero without any loss of generality. The final result reads:
\begin{widetext}
\begin{align}
\label{eq:alpha}
\pert{i}{\alpha}_{\ell} &= -\frac{2 \fsc \left( r^2\left( \fsc^{-1} \pert{i}{S}_{\ell\,\,00} + \fsc \pert{i}{S}_{\ell\,\,11} \right) + 2 \pert{i}{S}_{\ell\,\,+} \right)}{\ell(\ell+1)\left(\ell(\ell+1) - 2 \fsc + r \fsc'\right)} \, ,
\\
\label{eq:beta}
\pert{i}{\beta}_{\ell} &= \frac{1}{A} \left(r \partial_r \pert{i}{\alpha}_{\ell} - \frac{\ell(\ell+1) - 2 \fsc + r \fsc'}{2 \fsc} \pert{i}{\alpha}_{\ell}\right) \, ,
\\
\label{eq:gamma}
\pert{i}{\gamma}_{\ell} &= -\frac{ 2 r \left( r^2 \left( \fsc^{-1} \partial_t \pert{i}{S}_{\ell\,\,00} + \fsc  \partial_t \pert{i}{S}_{\ell\,\,11} \right) + 2  \partial_t \pert{i}{S}_{\ell\,\,+} - r \left( \ell(\ell+1) - 2 \fsc + r \fsc'\right) \pert{i}{S}_{\ell\,\,01} \right)}{\ell(\ell+1)\left(\ell(\ell+1) - 2 \fsc + r \fsc'\right)} \, .
\end{align}
\end{widetext}
\textit{Remark.} We remark that for maximally symmetric zero-order solution (M=0 in (\ref{eq:fsc_static})) $2\fsc - r \fsc' - 2 = 0$ holds and eqs. (\ref{eq:potential_static_polar}-\ref{eq:gamma}) simplify considerably; they are equivalent to analogous equations in \cite{r2017} (note the differences in definitions of the source functions between the present paper, eqs. (\ref{eq:f+}-\ref{eq:f01}), and eqs. (47-49) in \cite{r2017}), obtained in less general approach, well adapted to perturbations of maximally symmetric spacetimes in static coordinates (\ref{eq:line_element_static}). 

To summarize: to satisfy the set of perturbative Einstein equations (\ref{eq:pertEq}) for polar type perturbations of the zero-order solution (\ref{eq:line_element_static}) at axial symmetry, at any nonlinear order, for angular momenta $\ell \geq 2$ (cf. (\ref{eq:polarT_ab}-\ref{eq:polarT-})), it is enough to solve just one inhomogeneous wave equation for scalar master variable for polar perturbations (\ref{eq:scalar_wave}) with the potential and the scalar source given in (\ref{eq:potential_static_polar}) and (\ref{eq:source_scalar_wave}) respectively, and then reconstruct the Regge-Wheeler gauge invariant variables $\pert{i}{f}_{\ell\,\,00}$, $\pert{i}{f}_{\ell\,\,11}$, $\pert{i}{f}_{\ell\,\,01}$ and $\pert{i}{f}_{\ell\,\,+}$ according to (\ref{eq:f+}-\ref{eq:f01},\ref{eq:algebraicE-}) together with (\ref{eq:alpha}-\ref{eq:gamma}). Then, with a suitable gauge transformation, the resulting perturbations $\pert{i}{h}_{\ell\,\,\mu\nu}$ (cf. (\ref{eq:h00}-\ref{eq:h-})) can be put in asymptotically desired form \cite{r2017}.

The special cases $\ell=0$ and $\ell=1$ need a special treatment at nonlinear orders along the lines described in \cite{r2017}. 
%%%%%%%%%%%%%%%%%%%%%%%%%%%%%%%%%%%%%%%%%%%%%%%%%%%%%%%%%%%%%%%%%%%%%%%%%%%%%%
%%%%%%%%%%%%%%%%%%%%%%%%%%%%%%%%%%%%%%%%%%%%%%%%%%%%%%%%%%%%%%%%%%%%%%%%%%%%%%
%%%%%%%%%%%%%%%%%%%%%%%%%%%%%%%%%%%%%%%%%%%%%%%%%%%%%%%%%%%%%%%%%%%%%%%%%%%%%%

%%%%%%%%%%%%%%%%%%%%%%%%%%%%%%%%%%%%%%%%%%%%%%%%%%%%%%%%%%%%%%%%%%%%%%%%%%%%%%
%%%%%%%%%%%%%%%%%%%%%%%%%%%%%%%%%%%%%%%%%%%%%%%%%%%%%%%%%%%%%%%%%%%%%%%%%%%%%%
%%%%%%%%%%%%%%%%%%%%%%%%%%%%%%%%%%%%%%%%%%%%%%%%%%%%%%%%%%%%%%%%%%%%%%%%%%%%%%
\section{Perturbations of de Sitter solution in comoving coordinates}
\label{Sec:deSitter}
%%%%%%%%%%%%%%%%%%%%%%%%%%%%%%%%%%%%%%%%%%%%%%%%%%%%%%%%%%%%%%%%%%%%%%%%%%%%%%
%%%%%%%%%%%%%%%%%%%%%%%%%%%%%%%%%%%%%%%%%%%%%%%%%%%%%%%%%%%%%%%%%%%%%%%%%%%%%%
%%%%%%%%%%%%%%%%%%%%%%%%%%%%%%%%%%%%%%%%%%%%%%%%%%%%%%%%%%%%%%%%%%%%%%%%%%%%%%
To show that our scheme does not rely on the assumption of the zero-order solution being static, we apply it to the gravitational perturbations of de Sitter spacetime in comoving coordinates,
\begin{equation}
\label{eq:line_element_comoving}
ds^2 = -dt^2 + e^{2 t/\lsc} \left( dr^2 + r^2 d\Omega^2_2\right) 
\end{equation}
(to the best of our knowledge, the master scalar wave equation for polar perturbations of this solution has not been found, even at the linear order \cite{Viaggiu}). Our discussion will follow the lines of two previous sections. 

In axial symmetry, for polar perturbations of (\ref{eq:line_element_comoving}), we have (cf. (\ref{eq:hPolar},\ref{eq:polarT_ab}-\ref{eq:polarT-}))
\begin{align}
\label{eq:h00comoving}
\pert{i}{h}_{\ell\,\,00} = & \pert{i}{f}_{\ell\,\,00} + 2 \partial_t \pert{i}{\zeta}_{\ell\,\,0}\, ,
\\
\pert{i}{h}_{\ell\,\,11} = & \pert{i}{f}_{\ell\,\,11} - \frac{2 e^{2 t / \lsc}}{\lsc} \pert{i}{\zeta}_{\ell\,\,0} + 2 \partial_r \pert{i}{\zeta}_{\ell\,\,1} \, ,
\\
\pert{i}{h}_{\ell\,\,01} = & \pert{i}{f}_{\ell\,\,01} - \frac{2}{\lsc} \pert{i}{\zeta}_{\ell\,\,1} + \partial_r \pert{i}{\zeta}_{\ell\,\,0} + \partial_t \pert{i}{\zeta}_{\ell\,\,1} \, ,
\\
\pert{i}{h}_{\ell\,\,02} = & \pert{i}{\zeta}_{\ell\,\,0} + \partial_t \pert{i}{\zeta}_{\ell\,\,2} - \frac{2}{\lsc} \pert{i}{\zeta}_{\ell\,\,2} \, ,
\\
\pert{i}{h}_{\ell\,\,12} = & \pert{i}{\zeta}_{\ell\,\,1} - \frac{2}{r}\pert{i}{\zeta}_{\ell\,\,2} + \partial_r \pert{i}{\zeta}_{\ell\,\,2} \, ,
\\
\pert{i}{h}_{\ell\,\,+}  = & r^2 \pert{i}{f}_{\ell\,\,+} + 2 r \fsc \, \pert{i}{\zeta}_{\ell\,\,1} - \frac{2 e^{2 t / \lsc} r^2}{\lsc} \pert{i}{\zeta}_{\ell\,\,0} - \ell(\ell+1) \pert{i}{\zeta}_{\ell\,\,2} \, ,
\\
\label{eq:h-comoving}
\pert{i}{h}_{\ell\,\,-} = & \pert{i}{\zeta}_{\ell\,\,2} \, , 
\end{align}
where $\pert{i}{\zeta}_{\ell\,\,0}$, $\pert{i}{\zeta}_{\ell\,\,1}$, $\pert{i}{\zeta}_{\ell\,\,2}$ polar components define the $i$-th order polar gauge vector $\pert{i}{\zeta}_{\mu}$ (cf. (\ref{eq:polarV_a},\ref{eq:polarV_2})) and $\pert{i}{f}_{\ell\,\,00}(t,r)$, $\pert{i}{f}_{\ell\,\,11}(t,r)$, $\pert{i}{f}_{\ell\,\,01}(t,r)$, $\pert{i}{f}_{\ell\,\,+}(t,r)$ are  Regge-Wheeler type variables, being gauge invariant with respect to gauge transformations induced by $\pert{j}{\zeta}_{\mu}$ with $j \geq i$. Again, it is crucial to note that the sources in the polar sector of perturbative Einstein equations are not independent but fulfill three identities:
\begin{widetext}
\begin{align}
\label{eq:zero0comoving}
\frac{1}{2} \left( e^{2t / \lsc} \partial_t \pert{i}{S}_{\ell\,\,00} + \partial_t \pert{i}{S}_{\ell\,\,11}\right) + \frac{1}{r^2}\partial_t \pert{i}{S}_{\ell\,\,+} - \partial_r \pert{i}{S}_{\ell\,\,01} + \frac{3 e^{2t / \lsc}}{\lsc} \pert{i}{S}_{\ell\,\,00} - \frac{2}{r} \pert{i}{S}_{\ell\,\,01} + \frac{\ell(\ell+1)}{r^2} \pert{i}{S}_{\ell\,\,02} = 0\, ,
\\
\label{eq:zero1comoving}
\frac{1}{2} \left( e^{2t / \lsc} \partial_r \pert{i}{S}_{\ell\,\,00} + \partial_r \pert{i}{S}_{\ell\,\,11} \right) - \frac{1}{r^2}\partial_r \pert{i}{S}_{\ell\,\,+} - e^{2t / \lsc} \partial_t \pert{i}{S}_{\ell\,\,01} - \frac{3 e^{2t / \lsc}}{\lsc} \pert{i}{S}_{\ell\,\,01} + \frac{2}{r} \pert{i}{S}_{\ell\,\,11} - \frac{\ell(\ell+1)}{r^2} \pert{i}{S}_{\ell\,\,12} = 0\, ,
\\
\label{eq:zero2comoving}
\frac{1}{2} \left( e^{2t / \lsc} \pert{i}{S}_{\ell\,\,00} - \pert{i}{S}_{\ell\,\,11}\right) - e^{2t / \lsc} \partial_t \pert{i}{S}_{\ell\,\,02} + \partial_r \pert{i}{S}_{\ell\,\,12} + \frac{2}{r} \pert{i}{S}_{\ell\,\,12} - \frac{(\ell-1)(\ell+2)}{r^2} \pert{i}{S}_{\ell\,\,-} - \frac{3 e^{2t / \lsc}}{\lsc} \pert{i}{S}_{\ell\,\,02} = 0\, .
\end{align}
\end{widetext}
In analogy to (\ref{eq:zero0}-\ref{eq:zero2}), these identities follow from (three) polar components of $\bar \nabla^{\mu} \pert{i}{E}_{\mu\nu}=0$.

The polar components of the system of perturbative Einstein equations (\ref{eq:pertEq}) are expanded according to (\ref{eq:polarT_ab}-\ref{eq:polarT-}). We note that
\begin{equation}
\label{eq:algebraicE-comoving}
\pert{i}{E}_{\ell\,\,-} = \frac{1}{4} \left( \pert{i}{f}_{\ell\,\,00}  - e^{-2t / \lsc} \pert{i}{f}_{\ell\,\,11} \right) - \pert{i}{S}_{\ell\,\,-} = 0
\end{equation}
sets purely algebraic relation between $\pert{i}{f}_{\ell\,\,00}$ and
$\pert{i}{f}_{\ell\,\,11}$, while the combination
\begin{widetext}
\begin{align}
0 =& \frac{1}{2} \left( e^{2t / \lsc} \pert{i}{E}_{\ell\,\,00} - \pert{i}{E}_{\ell\,\,11} \right) - e^{2t / \lsc} \partial_t \pert{i}{E}_{\ell\,\,02}  + \partial_r \pert{i}{E}_{\ell\,\,12} - \frac{3 e^{2t / \lsc}}{\lsc} \pert{i}{E}_{\ell\,\,02} + \frac{2}{r} \pert{i}{E}_{\ell\,\,12} - \frac{(\ell-1)(\ell+2)}{r^2} \pert{i}{E}_{\ell\,\,-}  
\nonumber\\
\label{eq:sourceIdentitycomoving}
\equiv & \frac{1}{2} \left( e^{2t / \lsc} \pert{i}{S}_{\ell\,\,00} - \pert{i}{S}_{\ell\,\,11}\right) - e^{2t / \lsc} \partial_t \pert{i}{S}_{\ell\,\,02} + \partial_r \pert{i}{S}_{\ell\,\,12} - \frac{3 e^{2t / \lsc}}{\lsc} \pert{i}{S}_{\ell\,\,02} + \frac{2}{r} \pert{i}{S}_{\ell\,\,12} - \frac{(\ell-1)(\ell+2)}{r^2} \pert{i}{S}_{\ell\,\,-}
\end{align}
\end{widetext}
reduces to the identity (\ref{eq:zero2comoving}) fulfilled by the sources. Thus after eliminating $\pert{i}{f}_{\ell\,\,00}$ from (\ref{eq:algebraicE-comoving}) we are left with three RW gauge invariant variables $\pert{i}{f}_{\ell\,\,11}$, $\pert{i}{f}_{\ell\,\,01}$, $\pert{i}{f}_{\ell\,\,+}$ to satisfy five linearly independent equations: $\pert{i}{E}_{\ell\,\,00} = \pert{i}{E}_{\ell\,\,01} = \pert{i}{E}_{\ell\,\,+} = \pert{i}{E}_{\ell\,\,02} = \pert{i}{E}_{\ell\,\,12} = 0$.

We show again that we can easily fulfill all perturbative Einstein equations (\ref{eq:pertEq}) introducing master scalar variables along the lines of Sec.\ref{Sec:Intro}. First, we assume that these master scalar variables $\pert{i}{\Phi}_{\ell}^{\Psc} \equiv \pert{i}{\Phi}_{\ell}^{\Psc} (t,r)$ fulfill inhomogeneous (at nonlinear orders) linear scalar wave equation with a potential (\ref{eq:scalar_wave}), where now $\bar \Box$ is the scalar wave operator for the metric defined by the line element (\ref{eq:line_element_comoving}), and the potentials $V_{\ell}^{\Psc}$ and inhomogenities (scalar sources) $\pert{i}{\tilde S}_{\ell}^{\Psc}$ will be defined below. It is easy to check that for the line element
(\ref{eq:line_element_comoving})
\begin{equation}
\label{eq:scalar_wave_comoving}
\tilde \Box_{\ell}^{\Psc} \pert{i}{\Phi}_{\ell}^{\Psc} = \left(\partial_{tt} + \frac{3}{\lsc} \partial_t - e^{-2t / \lsc} \partial_{rr} + V_{\ell}^{\Psc} \right) \pert{i}{\Phi}_{\ell}^{\Psc}\, .
\end{equation}
Next, we express RW variables $\pert{i}{f}_{\ell\,\,11}$, $\pert{i}{f}_{\ell\,\,01}$ and $\pert{i}{f}_{\ell\,\,+}$ as linear combinations of master scalar variables $\pert{i}{\Phi}_{\ell}^{\Psc}$ and their (partial) derivatives up to the second order, and source functions to be defined below. The potential $V_{\ell}^{\Psc}$ in (\ref{eq:scalar_wave}) and linear combination of master scalar variables and their partial derivatives are chosen to satisfy satisfy the homogeneous part of perturbative Einstein equations (\ref{eq:pertEq}), while the source functions are chosen to satisfy inhomogenities of these equations. Indeed, there is a unique choice for the potential $V_{\ell}^{\Psc}$ in (\ref{eq:scalar_wave}), and a unique linear combination of master scalar variables (up to a multiplicative factor) that solves homogeneous part of (\ref{eq:pertEq}) identically, namely $\Delta_L \pert{i}{h}_{\mu \nu}
\equiv 0$ if and only if
\begin{equation}
\label{eq:potential_comoving_polar}
V_{\ell}^{\Psc} = \frac{2}{\lsc^2} + \frac{\ell(\ell+1)}{r^2} e^{-2t / \lsc} \, ,
\end{equation}
\begin{widetext}
\begin{align}
\label{eq:f+comoving}
\pert{i}{f}_{\ell\,\,+} = & e^{2t / \lsc} \left( - \frac{r}{\lsc} e^{2t / \lsc} \partial_{t} \pert{i}{\Phi}^{\Psc}_{\ell} + \partial_{r} \pert{i}{\Phi}^{\Psc}_{\ell} + \frac{1}{r} \left( \frac{\ell(\ell+1)}{2} - e^{2t / \lsc} \frac{r^2}{\lsc^2} \fsc \right) \pert{i}{\Phi}_{\ell}^{\Psc} \right) + \pert{i}{\alpha}_{\ell}\, ,
\\
\label{eq:f11comoving}
\pert{i}{f}_{\ell\,\,11} = & e^{2t / \lsc} \left( r \partial_{rr} \pert{i}{\Phi}_{\ell}^{\Psc} + \partial_{r} \pert{i}{\Phi}_{\ell}^{\Psc} - \frac{\ell(\ell+1)}{2 r} \pert{i}{\Phi}_{\ell}^{\Psc} \right) + \pert{i}{\beta}_{\ell}  
\\
\label{eq:f01comoving}
\pert{i}{f}_{\ell\,\,01} = & e^{2t / \lsc} \left( r \partial_{tr} \pert{1}{\Phi}^{\Psc}_{\ell} + \partial_{t} \pert{i}{\Phi}_{\ell}^{\Psc} + \frac{r}{\lsc} \partial_{r} \pert{i}{\Phi}_{\ell}^{\Psc} + \frac{1}{\lsc} \pert{i}{\Phi}_{\ell}^{\Psc} \right)  + \pert{i}{\gamma}_{\ell}\, ,
\end{align}
\end{widetext}
and we set the source functions in (\ref{eq:f+comoving}-\ref{eq:f01comoving}) and the scalar source in (\ref{eq:scalar_wave}) to zero: $\pert{i}{\alpha}_{\ell} =
\pert{i}{\beta}_{\ell} = \pert{i}{\gamma}_{\ell} = 0 = \pert{i}{\tilde S}_{\ell}^{\Psc}$ (their nonzero values will take care for the inhomogenities in perturbative Einstein equations (\ref{eq:pertEq})). We used here the scalar wave equation (\ref{eq:scalar_wave}) to eliminate from our expressions all partial derivatives of $\pert{i}{\Phi}_{\ell}^{\Psc}$ with respect to time, higher then the first order, and $\pert{i}{f}_{\ell\,\,00}$ is fixed from (\ref{eq:algebraicE-comoving})). In other words, eqs. (\ref{eq:algebraicE-comoving},\ref{eq:scalar_wave},\ref{eq:potential_comoving_polar}-\ref{eq:f01comoving}) constitute the solution of perturbative Einstein equations (\ref{eq:pertEq}) at the linear order $(i=1)$. Now, the scalar sources $\pert{i}{\tilde S}_{\ell}^{\Psc}$ and source functions $\pert{i}{\alpha}_{\ell}$, $\pert{i}{\beta}_{\ell}$, $\pert{i}{\gamma}_{\ell}$ can be set to fulfill perturbative Einstein equations at any nonlinear order. Again, to find $\pert{i}{\tilde S}_{\ell}^{\Psc}$, the relations (\ref{eq:f+comoving}-\ref{eq:f01comoving}) have to be inverted for $\pert{i}{\Phi}_{\ell}^{\Psc}$ itself. Moreover, such inversion is necessary to provide initial data for the scalar wave equation (\ref{eq:scalar_wave}) once the metric perturbations are prescribed at some initial time slice $t=\,$const. In general, such inversion is not uniquely defined, however it becomes uniquely defined if we wish to provide the initial data for $\pert{i}{\Phi}_{\ell}^{\Psc}$ at some initial time-slice in terms of the first and the second fundamental forms of an initial data hypersurface: 
\begin{widetext}
\begin{align}
\pert{i}{\Phi}^{\Psc}_{\ell} = & \frac{2 r^2}{(\ell-1)\ell(\ell+1)(\ell+2)} 
\left( e^{-2t / \lsc} r \partial_{tt} \pert{i}{f}_{\ell\,\,+} + \frac{r}{\lsc} \left( \partial_t \pert{i}{f}_{\ell\,\,11} + 2 \partial_t \pert{i}{f}_{\ell\,\,+} \right) - e^{-2t / \lsc} \left( \partial_r \pert{i}{f}_{\ell\,\,11} - \partial_r \pert{i}{f}_{\ell\,\,+}  \right) \right. 
\nonumber\\
& \hskip 3.5cm + \left. \left( \frac{r}{\lsc^2} - \frac{(\ell-1)(\ell+2)}{2r} \right) \left( \pert{i}{f}_{\ell\,\,11} - \pert{i}{f}_{\ell\,\,+}\right) - \frac{3 r}{\lsc^2} \pert{i}{f}_{\ell\,\,+} \right) \,,
\label{eq:Phi_def_comoving}
\end{align}
We note that the formula (\ref{eq:Phi_def_comoving}) contains first time derivatives, so it can not be used to provide the initial data for $\partial_t \pert{i}{\Phi}_{\ell}^{\Psc}$ at some initial time-slice in terms of the first and the second fundamental forms of an initial data hypersurface. To circumvent this problem we use different formula for $\partial_t \pert{1}{\Phi}_{\ell}^{\Psc}$ at linear order:
\begin{align}
\partial_t \pert{1}{\Phi}^{\Psc}_{\ell} = &\frac{e^{-4t / \lsc}}{(\ell-1)\ell(\ell+1)(\ell+2)} 
\nonumber\\
\times & \left( 
\frac{r \lsc}{2} \left( \ell(\ell+1) - 4 \frac{r^2}{\lsc^2} e^{2t / \lsc} \right) \partial_{rr} \pert{1}{f}_{\ell\,\,+} 
\right. 
\nonumber\\
& + \frac{r}{2} \left( \ell(\ell+1) - 4 \frac{r^2}{\lsc^2} e^{2t / \lsc} \right) e^{2t / \lsc} \partial_t \pert{1}{f}_{\ell\,\,11}
+ 2 r \left( \ell(\ell+1) - 2 \frac{r^2}{\lsc^2} e^{2t / \lsc} \right) e^{2t / \lsc} \partial_t \pert{1}{f}_{\ell\,\,+}
\nonumber\\
& - \frac{\lsc}{2} \left( \ell(\ell+1) - 4 \frac{r^2}{\lsc^2} e^{2t / \lsc} \right) \partial_r \pert{1}{f}_{\ell\,\,11}
+ \frac{\lsc}{2} \left( 3 \ell(\ell+1) - 4 \frac{r^2}{\lsc^2} e^{2t / \lsc} \right) \partial_r \pert{1}{f}_{\ell\,\,+}
\nonumber\\
& - \frac{\lsc}{4r} \left( \ell(\ell+1)( \ell(\ell+1) + 2 ) + 2 \left( 4 - 5 \ell(\ell+1) + 4 \frac{r^2}{\lsc^2} e^{2t / \lsc} \right)\frac{r^2}{\lsc^2} e^{2t / \lsc} \right) \pert{1}{f}_{\ell\,\,11}
\nonumber\\
& \left. + \frac{\lsc}{4r} \left( -(\ell-1)\ell(\ell+1)(\ell+2) + 4 \left( 2 - 5 \ell(\ell+1) + 8 \frac{r^2}{\lsc^2} e^{2t / \lsc} \right)\frac{r^2}{\lsc^2} e^{2t / \lsc} \right) \pert{1}{f}_{\ell\,\,+} \right)
\label{eq:DotPhi_def_comoving}
\end{align}
%\end{widetext}
Indeed, at linear order, upon substituting (\ref{eq:f+comoving},\ref{eq:f11comoving}) with (\ref{eq:scalar_wave}) fulfilled (\ref{eq:Phi_def_comoving}) and (\ref{eq:DotPhi_def_comoving}) become identities. At nonlinear orders we take (\ref{eq:Phi_def_comoving}) as a \textit{definition} of master scalar variables for polar perturbations and discard (\ref{eq:DotPhi_def_comoving})
\footnote{We are allowed to do it because we take care for the initial
  metric perturbations through the initial data for the first order
  scalar wave equation ((\ref{eq:scalar_wave}) with $i=1$), and then
  the initial data for higher orders scalar wave equations
  ((\ref{eq:scalar_wave}) with $i>0$) are identically zero.}.
The definition (\ref{eq:Phi_def_comoving}) fixes the scalar source $\pert{i}{\tilde S}_{\ell}^{\Psc}$ in the inhomogeneous scalar wave equation (\ref{eq:scalar_wave}): 
%begin{widetext}
\begin{align}
\label{eq:source_scalar_wave_comoving}
& \pert{i}{\tilde S}^{\Psc}_{\ell} = \frac {2 r} {(\ell-1)\ell(\ell+1)(\ell+2)} \times 
\nonumber\\
& \left( 
r^2 \partial_{rr} 
\left( \pert{i}{S}_{\ell\,\,00} + e^{-2t / \lsc} \pert{i}{S}_{\ell\,\,11} \right) 
- 2 r^2 \partial_{tr} \pert{i}{S}_{\ell\,\,01} 
- \frac{r^2}{\lsc} 
\partial_t \left( e^{2t / \lsc} \pert{i}{S}_{\ell\,\,00} + \pert{i}{S}_{\ell\,\,11} \right) 
+ 4 r \partial_r 
\left( \pert{i}{S}_{\ell\,\,00} + e^{-2t / \lsc} \pert{i}{S}_{\ell\,\,11} \right) 
\right.
\nonumber\\ & \hskip 2mm 
+ 2r 
\left( 
- \partial_t \pert{i}{S}_{\ell\,\,01} 
- \frac{r}{\lsc} \partial_r \pert{i}{S}_{\ell\,\,01} 
- \frac{r}{\lsc} e^{2t / \lsc} \partial_{tt} \pert{i}{S}_{\ell\,\,02} 
- \partial_{tr} \pert{i}{S}_{\ell\,\,02}
+ 2 \frac{r}{\lsc} \partial_{rr} \pert{i}{S}_{\ell\,\,02}
- \frac{r}{\lsc} \partial_{tr} \pert{i}{S}_{\ell\,\,12}
+ e^{-2t / \lsc} \partial_{rr} \pert{i}{S}_{\ell\,\,12} 
\right)
\nonumber\\ & \hskip 2mm 
+ \left( \frac{2}{\lsc^2} + \frac{\ell(\ell+1)}{r^2} e^{-2t / \lsc} \right) \pert{i}{S}_{\ell\,\,+}
- \left( 2 \frac{r^2}{\lsc^2} + \frac{\ell(\ell+1)-4}{2} e^{-2t / \lsc}\right) \left( e^{2t / \lsc} \pert{i}{S}_{\ell\,\,00} + \pert{i}{S}_{\ell\,\,11} \right)
+  2 \frac{r}{\lsc} \pert{i}{S}_{\ell\,\,01}
\nonumber\\ & \hskip 2mm 
+ \left( -16 \frac{r^2}{\lsc^2} e^{2t / \lsc} + (\ell-1)(\ell+2) \right) \partial_t \pert{i}{S}_{\ell\,\,02}
+ 18 \frac{r}{\lsc} \partial_r \pert{i}{S}_{\ell\,\,02}
-  4 \frac{r}{\lsc} \partial_t \pert{i}{S}_{\ell\,\,12}
- \left( 2 \frac{r^2}{\lsc^2} + (\ell-2)(\ell+3) e^{-2t / \lsc} \right) \partial_r \pert{i}{S}_{\ell\,\,12}
\nonumber\\ & \hskip 2mm 
+ 12 \frac{r^2}{\lsc^2} \left( -e^{2t / \lsc} \partial_{tt} \pert{i}{S}_{\ell\,\,-} + \partial_{rr} \pert{i}{S}_{\ell\,\,-} \right)
- \frac{1}{\lsc} \left( 30 \frac{r^2}{\lsc^2} e^{2t / \lsc} + \ell(\ell+1) - 18 \right) \pert{i}{S}_{\ell\,\,02}
- \frac{2}{r} \left( 2 \frac{r^2}{\lsc^2} + \ell(\ell+1) e^{-2t / \lsc} \right) \pert{i}{S}_{\ell\,\,12}
\nonumber\\ & \hskip 2mm 
+ \frac{2}{\lsc} \left( -42 \frac{r^2}{\lsc^2} e^{2t / \lsc} + (\ell-1)(\ell+2) \right) \partial_t \pert{i}{S}_{\ell\,\,-}
+ \frac{2}{r} \left( 36 \frac{r^2}{\lsc^2} e^{2t / \lsc} - (\ell-1)(\ell+2) \right) \partial_r \pert{i}{S}_{\ell\,\,-} 
\nonumber\\ & \hskip 2mm 
\left. - \frac{2}{\lsc^2} \left( 72 \frac{r^2}{\lsc^2} e^{2t / \lsc} + 5 \ell(\ell+1) -34 - (\ell-1)(\ell+2)(\ell^2 + \ell + 1) \frac{\lsc^2}{r^2} e^{-2t / \lsc} \right) \pert{i}{S}_{\ell\,\,-} \right)
\end{align}
\end{widetext}
for $\ell \geq 2$ (here the $\ell=0$ and $\ell=1$ cases have to be treated separately). Again, we note that, the formula (\ref{eq:source_scalar_wave_comoving}) is unique up to linear combination of the identities (\ref{eq:zero0comoving}-\ref{eq:zero2comoving}) and their derivatives. Finally, we are ready to set the source functions $\pert{i}{\alpha}_{\ell}$, $\pert{i}{\beta}_{\ell}$ and $\pert{i}{\gamma}_{\ell}$ in (\ref{eq:f+comoving}-\ref{eq:f01comoving}). They are set in such a way that all perturbative Einstein equations (\ref{eq:pertEq}) are satisfied and (\ref{eq:Phi_def_comoving}) becomes the identity upon substituting (\ref{eq:f+comoving},\ref{eq:f11comoving}) with (\ref{eq:scalar_wave}) fulfilled, also at nonlinear order. It is crucial to note, that without identities (\ref{eq:zero0comoving}-\ref{eq:zero2comoving}) it would be impossible to meet all these nonlinear requirements simultaneously. Moreover it is worth to note that due to the identities (\ref{eq:zero0comoving}-\ref{eq:zero2comoving}) the solutions for $\pert{i}{\alpha}_{\ell}$, $\pert{i}{\beta}_{\ell}$ and $\pert{i}{\gamma}_{\ell}$ can be found in purely algebraic way, without solving any (partial) differential equations. Namely, we write down the source functions $\pert{i}{\alpha}_{\ell}$, $\pert{i}{\beta}_{\ell}$ and $\pert{i}{\gamma}_{\ell}$ as linear combination of the sources $\pert{i}{S}_{\ell\,\,\mu\nu}$ and their first, and second derivatives. Fixing $3 \times 7 \times 6 = 126$ function coefficients of these linear combinations is a technical task with \textit{Mathematica}. It turns out that, out of these $126$ functions, $99$ functions are uniquely fixed in terms of $27$ free functions, and moreover, in the resulting expressions, the coefficients of $27$ free functions are identically zero, due to the identities (\ref{eq:zero0comoving}-\ref{eq:zero2comoving}). Thus we can put them to zero without any loss of generality. The final result read
%The final expressions are quite lengthy, so we give here explicitly the result for $\pert{i}{\alpha}_{\ell}$ and $\pert{i}{\beta}_{\ell}$ only:
\begin{widetext}
\begin{align}
\label{eq:alpha_comoving}
& (\ell-1)^2\ell(\ell+1)(\ell+2)^2 e^{-4t / \lsc} \,\, \pert{i}{\alpha}_{\ell} 
\nonumber\\ 
= & \frac{24 r^5}{\lsc^2} \partial_{rr} \pert{i}{S}_{\ell\,\,12}  
- \frac{24 e^{2 t / \lsc} r^5}{\lsc^2} \partial_{tr} \pert{i}{S}_{\ell\,\,02} 
- \frac{\left(7 (\ell-1) (\ell+2) \lsc^2+36 e^{2 t / \lsc} r^2\right) r^4}{\lsc^3} \partial_r \pert{i}{S}_{\ell\,\,01}  
- \frac{24 e^{2 t / \lsc} (\ell-1) (\ell+2) r^4}{\lsc^3} \partial_t \pert{i}{S}_{\ell\,\,-} 
\nonumber\\ 
+ & \frac{e^{2 t / \lsc} \left(5 (\ell-1) (\ell+2) \lsc^2 + 36 e^{2 t / \lsc} r^2\right) r^4}{2 \lsc^3} \partial_t \pert{i}{S}_{\ell\,\,00} 
- \frac{8 e^{2 t / \lsc} \left(\ell^2+\ell-8\right) r^4}{\lsc^2} \partial_t \pert{i}{S}_{\ell\,\,02} 
\nonumber\\ 
+ & \frac{8 \left((\ell-1) (\ell+2) \lsc^2-9 e^{2 t / \lsc} r^2\right) r^3}{\lsc^3} \partial_r \pert{i}{S}_{\ell\,\,02} 
+ \frac{e^{-2 t / \lsc} \left(6 e^{2 t / \lsc} \left(\ell^2+\ell-26\right) r^2-(\ell-2) (\ell-1) (\ell+2) (\ell+3) \lsc^2\right) r^3}{8 \lsc^2} \partial_r \pert{i}{S}_{\ell\,\,11} 
\nonumber\\ 
+ & \frac{\left(6 e^{2 t / \lsc} \left(3\ell^2+3 \ell+50\right) r^2-(\ell-1) (\ell+2) \left(3 \ell^2+3\ell-34\right) \lsc^2\right) r^3}{8 \lsc^2} \partial_t \pert{i}{S}_{\ell\,\,01} 
\nonumber\\ 
+ & \frac{\left((\ell-2) (\ell-1) (\ell+2) (\ell+3) \lsc^2-6 e^{2 t / \lsc} \left(\ell^2+\ell+6\right) r^2\right) r^3}{16 \lsc} \partial_{tr} \pert{i}{S}_{\ell\,\,00} 
\nonumber\\ 
- & \frac{e^{-2 t / \lsc} \left(6 e^{2 t / \lsc} \left(\ell^2+\ell+6\right) r^2-(\ell-2) (\ell-1) (\ell+2) (\ell+3) \lsc^2\right) r^3}{16 \lsc} \partial_{tr} \pert{i}{S}_{\ell\,\,11} 
\nonumber\\ 
- & \frac{\left((\ell-2) (\ell-1) (\ell+2) (\ell+3) \lsc^2-6 e^{2 t / \lsc} \left(\ell^2+\ell+6\right) r^2\right) r^3}{8 \lsc} \partial_{tt} \pert{i}{S}_{\ell\,\,01} 
\nonumber\\ 
+ & \frac{12 (\ell-1) (\ell+2) \left(\left(\ell^2+\ell+14\right) \lsc^2-6 e^{2 t / \lsc} r^2\right) r^2}{\lsc^4} \pert{i}{S}_{\ell\,\,-} 
\nonumber\\ 
+ & \frac{\left((\ell-1) (\ell+2) \left(\ell^2+\ell+2\right) \lsc^4+16 e^{2 t / \lsc} \left(2 \ell^2+2 \ell-7\right) r^2 \lsc^2+216 e^{4 t / \lsc} r^4\right) r^2}{2 \lsc^4} \pert{i}{S}_{\ell\,\,00} 
\nonumber\\ 
+ & \frac{\left((\ell-1) (\ell+2) \left(11 \ell^2+11 \ell+24\right) \lsc^2+12 e^{2 t / \lsc} \left(\ell^2+\ell+16\right) r^2\right) r^2}{\lsc^3} \pert{i}{S}_{\ell\,\,02} 
+ \frac{\left(5 (\ell-1) (\ell+2) \lsc^2+36 e^{2 t / \lsc} r^2\right) r^2}{\lsc^3} \partial_t \pert{i}{S}_{\ell\,\,+} 
\nonumber\\ 
+ & \frac{e^{-2 t / \lsc} \left((\ell-2) (\ell-1) (\ell+2) (\ell+3) \lsc^4+4 e^{2 t / \lsc} \left(\ell^2+\ell-14\right) r^2 \lsc^2+72 e^{4 t / \lsc} r^4\right) r^2}{4 \lsc^3} \partial_t \pert{i}{S}_{\ell\,\,11} 
\nonumber\\ 
- & \frac{e^{-2 t / \lsc} \left(6 e^{2 t / \lsc} \left(\ell^4+2 \ell^3-9 \ell^2-10 \ell+96\right) r^2-(\ell-1) \ell (\ell+1) (\ell+2) \left(\ell^2+\ell+2\right) \lsc^2\right) r}{4 \lsc^2} \pert{i}{S}_{\ell\,\,12} 
\nonumber\\ 
- &\frac{e^{-2 t / \lsc} \left(6 e^{2 t / \lsc} \left(\ell^2+\ell-10\right) r^2-(\ell-1) (\ell+2) \left(\ell^2+\ell+10\right) \lsc^2\right) r}{4 \lsc^2} \partial_r \pert{i}{S}_{\ell\,\,+} 
\nonumber\\ 
+ & \frac{e^{-2 t / \lsc} \ell (\ell+1) \left(6 e^{2 t / \lsc} \left(\ell^2+\ell+6\right) r^2-(\ell-2) (\ell-1) (\ell+2) (\ell+3) \lsc^2\right) r}{8 \lsc} \partial_t \pert{i}{S}_{\ell\,\,12} 
\nonumber\\ 
+ & \frac{e^{-2 t / \lsc} \left(6 e^{2 t / \lsc} \left(\ell^2+\ell+6\right) r^2-(\ell-2) (\ell-1) (\ell+2) (\ell+3) \lsc^2\right) r}{8 \lsc} \partial_{tr} \pert{i}{S}_{\ell\,\,+} 
\nonumber\\ 
+ & \frac{e^{-2 t / \lsc} (\ell-1) (\ell+2) \left((\ell-1) (\ell+2) \lsc^2+2 e^{2 t / \lsc} r^2\right)}{\lsc^2} \pert{i}{S}_{\ell\,\,+} \, ,
\end{align}

\begin{align}
\label{eq:gamma_comoving}
& (\ell-1)^2\ell(\ell+1)(\ell+2)^2 e^{-4t / \lsc} \,\, \pert{i}{\gamma}_{\ell} 
\nonumber\\ 
= & \frac{288 e^{2 t / \lsc} r^7}{\ell(\ell+1) \lsc^3} \partial_{rr} \pert{i}{S}_{\ell\,\,00}
+ \frac{288 r^7}{\ell (\ell+1) \lsc^3} \partial_{rr} \pert{i}{S}_{\ell\,\,11} 
+ \frac{24 \left((\ell-1) (\ell+2) \lsc^2 - 9 e^{2 t / \lsc} r^2\right) r^6}{\ell (\ell+1) \lsc^4} \partial_{rr} \pert{i}{S}_{\ell\,\,01}
+ \frac{72 r^6}{\lsc^3} \partial_{rr} \pert{i}{S}_{\ell\,\,12} 
\nonumber\\ 
- & \frac{192 e^{2 t / \lsc} r^6}{\ell (\ell+1) \lsc^2} \partial_{tt} \pert{i}{S}_{\ell\,\,01} 
+ \frac{48 \left(4 (\ell-1) (\ell+2) \lsc^2 - 93 e^{2 t / \lsc} r^2\right) r^5}{\ell (\ell+1) \lsc^4} \partial_r \pert{i}{S}_{\ell\,\,01} 
- \frac{576 r^5}{\ell (\ell+1) \lsc^3} \partial_{rr} \pert{i}{S}_{\ell\,\,+} 
\nonumber\\ 
+ & \frac{16 \left(2 (\ell-1) (\ell+2) \lsc^2 - 93 e^{2 t / \lsc} r^2\right) r^5}{\ell (\ell+1) \lsc^3} \partial_{tr} \pert{i}{S}_{\ell\,\,01}
+ \frac{96 r^5}{\lsc^2} \partial_{tr} \pert{i}{S}_{\ell\,\,12} 
+ \frac{8 e^{2 t / \lsc} \left(57 e^{2 t / \lsc} r^2 - 2 (\ell-1) (\ell+2) \lsc^2 \right) r^5}{\ell (\ell+1) \lsc^3} \partial_{tt} \pert{i}{S}_{\ell\,\,00} 
\nonumber\\ 
- & \frac{96 e^{2 t / \lsc} r^5}{\lsc^2} \partial_{tt} \pert{i}{S}_{\ell\,\,02} 
- \frac{8 \left(2 (\ell-1) (\ell+2) \lsc^2 - 57 e^{2 t / \lsc} r^2\right) r^5}{\ell (\ell+1) \lsc^3} \partial_{tt} \pert{i}{S}_{\ell\,\,11} 
\nonumber\\ 
+ & \frac{\left(5 (\ell-1) \ell (\ell+1) (\ell+2) \lsc^4 - 12 e^{2 t / \lsc} \left(3 \ell^2 + 3 \ell - 106\right) r^2 \lsc^2+648 e^{4 t / \lsc} r^4\right) r^4}{\ell (\ell+1) \lsc^5} \partial_r \pert{i}{S}_{\ell\,\,00} 
\nonumber\\ 
- & \frac{3 e^{-2 t / \lsc} \left(12 e^{2 t / \lsc} \left(\ell^2 + \ell - 58\right) r^2 - (\ell-1) \ell (\ell+1) (\ell+2) \lsc^2\right) r^4}{\ell (\ell+1) \lsc^3} \partial_r \pert{i}{S}_{\ell\,\,11} 
\nonumber\\ 
+ & \frac{16 \left(4 (\ell-1) (\ell+2) \lsc^2-291 e^{2 t / \lsc} r^2\right) r^4}{\ell (\ell+1) \lsc^3} \partial_t \pert{i}{S}_{\ell\,\,01} 
+ \frac{24 (\ell-1) (\ell+2) r^4}{\lsc^2} \partial_{tr} \pert{i}{S}_{\ell\,\,-} 
\nonumber\\ 
+ & \frac{\left((\ell-1) \ell (\ell+1) (\ell+2) \lsc^4 - 12 e^{2 t / \lsc} \left(\ell^2 + \ell - 10\right) r^2 \lsc^2 + 108 e^{4 t / \lsc} r^4\right) r^4}{\ell (\ell+1) \lsc^4} \partial_{tr} \pert{i}{S}_{\ell\,\,00} 
\nonumber\\ 
+ & \frac{8 \left((\ell-1) (\ell+2) \lsc^2 - 9  e^{2 t / \lsc} r^2\right) r^4}{\lsc^3} \partial_{tr} \pert{i}{S}_{\ell\,\,02} 
\nonumber\\ 
+ & \frac{e^{-2 t / \lsc} \left((\ell-1) \ell (\ell+1) (\ell+2) \lsc^4 - 12 e^{2 t / \lsc} \left(\ell^2 + \ell - 10\right) r^2 \lsc^2 + 108 e^{4 t / \lsc} r^4\right) r^4}{\ell (\ell+1) \lsc^4} \partial_{tr} \pert{i}{S}_{\ell\,\,11} 
\nonumber\\ 
+ & \frac{8 \left((\ell-1) \ell (\ell+1) (\ell+2) \lsc^4 - 39 e^{2 t / \lsc} \left(\ell^2 + \ell - 4\right) r^2 \lsc^2 + 1548 e^{4 t / \lsc} r^4\right) r^3}{\ell (\ell+1) \lsc^5} \pert{i}{S}_{\ell\,\,00} 
\nonumber\\ 
- & \frac{12 e^{-2 t / \lsc} \left(18 e^{2 t / \lsc} (\ell-3) (\ell+4) r^2 - (\ell-1) \ell (\ell+1) (\ell+2) \lsc^2\right) r^3}{\ell (\ell+1) \lsc^3} \pert{i}{S}_{\ell\,\,11}
- \frac{48 \left(2 (\ell-1) (\ell+2) \lsc^2 - 39 e^{2 t / \lsc} r^2\right) r^3}{\ell (\ell+1) \lsc^4} \partial_t \pert{i}{S}_{\ell\,\,+} 
\nonumber\\ 
+ & \frac{2 \left((\ell-1) \ell (\ell+1) (\ell+2) \lsc^4 - 4 e^{2 t / \lsc} \left(19 \ell^2 + 19 \ell - 50\right) r^2 \lsc^2 + 2400 e^{4 t / \lsc} r^4\right) r^3}{\ell (\ell+1) \lsc^4} \partial_t \pert{i}{S}_{\ell\,\,00} 
\nonumber\\ 
+ & \frac{2 e^{-2 t / \lsc} \left((\ell-1) \ell (\ell+1) (\ell+2) \lsc^4 - 12 e^{2 t / \lsc} \left(5 \ell^2 + 5 \ell - 22\right) r^2 \lsc^2 + 576 e^{4 t / \lsc} r^4\right) r^3}{\ell (\ell+1) \lsc^4} \partial_t \pert{i}{S}_{\ell\,\,11} 
\nonumber\\ 
- & \frac{16 \left(2 (\ell-1) (\ell+2) \lsc^2 - 57 e^{2 t / \lsc} r^2\right) r^3}{\ell (\ell+1) \lsc^3} \partial_{tt} \pert{i}{S}_{\ell\,\,+} 
\nonumber\\ 
- & \frac{2 e^{-2 t / \lsc} \left(-(\ell-1)^2 \ell (\ell+1) (\ell+2)^2 \lsc^4 - 120 e^{2 t / \lsc} (\ell-1) (\ell+2) r^2 \lsc^2 + 5472 e^{4 t / \lsc} r^4\right) r^2}{\ell (\ell+1) \lsc^4} \pert{i}{S}_{\ell\,\,01} 
\nonumber\\ 
- & \frac{2 e^{-2 t / \lsc} \left((\ell-1) \ell (\ell+1) (\ell+2) \lsc^2 + 360 e^{2 t / \lsc} r^2\right) r^2}{\ell (\ell+1) \lsc^3} \partial_r \pert{i}{S}_{\ell\,\,+} 
\nonumber\\ 
+ & \frac{2 e^{-2 t / \lsc} \left((\ell-1) \ell (\ell+1) (\ell+2) \lsc^4 - 12 e^{2 t / \lsc} \left(\ell^2 + \ell + 6\right) r^2 \lsc^2 + 108 e^{4 t / \lsc} r^4\right) r^2}{\ell (\ell+1) \lsc^4} \partial_{tr} \pert{i}{S}_{\ell\,\,+} 
- \frac{4 e^{-2 t / \lsc} (\ell-1) (\ell+2) r}{\lsc} \pert{i}{S}_{\ell\,\,+} \, ,
\end{align}

\begin{align}
\label{eq:beta_comoving}
& (\ell-1)^2\ell(\ell+1)(\ell+2)^2 e^{-4t / \lsc} \,\, \pert{i}{\beta}_{\ell} 
\nonumber\\ 
= & \frac{168 r^5}{\lsc^2} \partial_{rr} \pert{i}{S}_{\ell\,\,12}  
- \frac{168 e^{2 t / \lsc} r^5}{\lsc^2} \partial_{tr} \pert{i}{S}_{\ell\,\,02} 
+ \frac{10 (\ell-1) (\ell+2) r^4}{\lsc} \partial_r \pert{i}{S}_{\ell\,\,01} 
+ \frac{24 (\ell-1) (\ell+2) r^4}{\lsc^2} \partial_{rr} \pert{i}{S}_{\ell\,\,-} 
\nonumber\\ 
+ & \frac{8 (\ell-1) (\ell+2) r^4}{\lsc} \partial_{rr} \pert{i}{S}_{\ell\,\,02} 
+ 2 (\ell-1) (\ell+2) r^4 \partial_{tr} \pert{i}{S}_{\ell\,\,01}  
+ \frac{2 \left(23 (\ell-1) (\ell+2) \lsc^2+252 e^{2 t / \lsc} r^2\right) r^3}{\lsc^3} \pert{i}{S}_{\ell\,\,01} 
\nonumber\\ 
+ & \frac{56 \left((\ell-1) (\ell+2) \lsc^2-9 e^{2 t / \lsc} r^2\right) r^3}{\lsc^3} \partial_r \pert{i}{S}_{\ell\,\,02} 
- \frac{3 e^{-2 t / \lsc} \left((\ell-1) (\ell+2) \left(\ell^2+\ell+2\right) \lsc^2+336 e^{2 t / \lsc} r^2\right) r^3}{8 \lsc^2} \partial_r \pert{i}{S}_{\ell\,\,11} 
\nonumber\\ 
+ & \frac{\left(2352 e^{2 t / \lsc} r^2-(\ell-1) (\ell+2) \left(9 \ell^2+9 \ell-158\right) \lsc^2\right) r^3}{8 \lsc^2} \partial_t \pert{i}{S}_{\ell\,\,01} 
\nonumber\\ 
+ & \frac{\left((\ell-1) (\ell+2) \left(3 \ell^2+3 \ell-26\right) \lsc^2-336 e^{2 t / \lsc} r^2\right) r^3}{16 \lsc} \partial_{tr} \pert{i}{S}_{\ell\,\,00} 
\nonumber\\ 
- & \frac{e^{-2 t / \lsc} \left(336 e^{2 t / \lsc} r^2-(\ell-1) (\ell+2) \left(3 \ell^2+3 \ell-26\right) \lsc^2\right) r^3}{16 \lsc} \partial_{tr} \pert{i}{S}_{\ell\,\,11} 
\nonumber\\ 
- & \frac{\left((\ell-1) (\ell+2) \left(3 \ell^2+3 \ell-26\right) \lsc^2-336 e^{2 t / \lsc} r^2\right) r^3}{8 \lsc} \partial_{tt} \pert{i}{S}_{\ell\,\,01} 
- \frac{\left((\ell-2) (\ell-1) (\ell+2) (\ell+3) \lsc^2+336 e^{2 t / \lsc} r^2\right) r^2}{2 \lsc^2} \pert{i}{S}_{\ell\,\,00} 
\nonumber\\ 
+ & \frac{6 \left((\ell-1) (\ell+2) \left(\ell^2+\ell+12\right) \lsc^2+168 e^{2 t / \lsc} r^2\right) r^2}{\lsc^3} \pert{i}{S}_{\ell\,\,02} 
+ 4 e^{-2 t / \lsc} (\ell-1) (\ell+2) r^2 \partial_{rr} \pert{i}{S}_{\ell\,\,+} 
\nonumber\\ 
+ & \frac{2 \left((\ell-1) \ell (\ell+1) (\ell+2) \lsc^2+168 e^{2 t / \lsc} r^2\right) r^2}{\lsc^2} \partial_t \pert{i}{S}_{\ell\,\,02} 
\nonumber\\ 
- & \frac{e^{-2 t / \lsc} \left(336 e^{2 t / \lsc} r^2-(\ell-1) (\ell+2) \left(3 \ell^2+3 \ell-26\right) \lsc^2\right) r^2}{4 \lsc} \partial_t \pert{i}{S}_{\ell\,\,11} 
\nonumber\\ 
+ & \frac{3 e^{-2 t / \lsc} \left((\ell-2) (\ell-1) \ell (\ell+1) (\ell+2) (\ell+3) \lsc^2+112 e^{2 t / \lsc} (\ell-3) (\ell+4) r^2\right) r}{4 \lsc^2} \pert{i}{S}_{\ell\,\,12} 
\nonumber\\ 
+ & \frac{3 e^{-2 t / \lsc} \left((\ell-1) (\ell+2) \left(\ell^2+\ell+2\right) \lsc^2+112 e^{2 t / \lsc} r^2\right) r}{4 \lsc^2} \partial_r \pert{i}{S}_{\ell\,\,+} 
\nonumber\\ 
+ & \frac{e^{-2 t / \lsc} \ell (\ell+1) \left(336 e^{2 t / \lsc} r^2-(\ell-1) (\ell+2) \left(3 \ell^2+3 \ell-26\right) \lsc^2\right) r}{8 \lsc} \partial_t \pert{i}{S}_{\ell\,\,12} 
\nonumber\\ 
+ & \frac{e^{-2 t / \lsc} \left(336 e^{2 t / \lsc} r^2-(\ell-1) (\ell+2) \left(3 \ell^2+3 \ell-26\right) \lsc^2\right) r}{8 \lsc} \partial_{tr} \pert{i}{S}_{\ell\,\,+} 
\nonumber\\ 
- & \frac{2  e^{-2 t / \lsc} (\ell-1) (\ell+2) \left(6 e^{2 t / \lsc} \left(\ell^2+\ell-74\right) r^2-(\ell-1) \ell (\ell+1) (\ell+2) \lsc^2\right)}{\lsc^2} \pert{i}{S}_{\ell\,\,-} 
+ e^{-2 t / \lsc} (\ell-1)^2 (\ell+2)^2 \pert{i}{S}_{\ell\,\,+} \, .
\end{align}
\end{widetext}
%%%%%%%%%%%%%%%%%%%%%%%%%%%%%%%%%%%%%%%%%%%%%%%%%%%%%%%%%%%%%%%%%%%%%%%%%%%%%%
%%%%%%%%%%%%%%%%%%%%%%%%%%%%%%%%%%%%%%%%%%%%%%%%%%%%%%%%%%%%%%%%%%%%%%%%%%%%%%
%%%%%%%%%%%%%%%%%%%%%%%%%%%%%%%%%%%%%%%%%%%%%%%%%%%%%%%%%%%%%%%%%%%%%%%%%%%%%%

%%%%%%%%%%%%%%%%%%%%%%%%%%%%%%%%%%%%%%%%%%%%%%%%%%%%%%%%%%%%%%%%%%%%%%%%%%%%%%
%%%%%%%%%%%%%%%%%%%%%%%%%%%%%%%%%%%%%%%%%%%%%%%%%%%%%%%%%%%%%%%%%%%%%%%%%%%%%%
%%%%%%%%%%%%%%%%%%%%%%%%%%%%%%%%%%%%%%%%%%%%%%%%%%%%%%%%%%%%%%%%%%%%%%%%%%%%%%
\section{Summary.}
\label{Sec:summary}
%%%%%%%%%%%%%%%%%%%%%%%%%%%%%%%%%%%%%%%%%%%%%%%%%%%%%%%%%%%%%%%%%%%%%%%%%%%%%%
%%%%%%%%%%%%%%%%%%%%%%%%%%%%%%%%%%%%%%%%%%%%%%%%%%%%%%%%%%%%%%%%%%%%%%%%%%%%%%
%%%%%%%%%%%%%%%%%%%%%%%%%%%%%%%%%%%%%%%%%%%%%%%%%%%%%%%%%%%%%%%%%%%%%%%%%%%%%%
We have presented a systematic and robust approach to nonlinear gravitational perturbations of vacuum spacetimes. In particular, we showed that the system of perturbative Einstein equations (\ref{eq:pertEq}) reduces at each perturbation order to two (for each gravitational mode in $3+1$ dimensions) scalar wave equations (\ref{eq:scalar_wave}), and then we showed how the metric perturbations can be explicitly obtained, once the solutions to these scalar wave equations are known, cf. (\ref{eq:f+}-\ref{eq:f01},\ref{eq:f+comoving}-\ref{eq:f01comoving}) together with (\ref{eq:alpha}-\ref{eq:gamma},\ref{eq:alpha_comoving}-\ref{eq:beta_comoving}). These scalar wave equations correspond to two polarization states of a gravitational wave, thus extending the concept of polarization beyond linear approximation. Our scheme provides a powerful tool to study gravitational radiation and black holes stability beyond the linear order. In the next steps we intend to focus on applications both in vacuum (nonlinear quasi-normal modes couplings and radiation from perturbed black holes) and with matter (self-force problems, accretion disks around black holes, extreme-mass-ratio inspirals and finally cosmological perturbations).

From \textit{a posteriori} perspective it is seen that the profound understanding of linear perturbations together with the identities fulfilled by the sources in perturbative Einstein equations (\ref{eq:pertEq}) at nonlinear orders, cf. (\ref{eq:zero0}-\ref{eq:zero2},\ref{eq:zero0comoving}-\ref{eq:zero2comoving}) are the keys to nonlinear gravitational perturbations. 

We used a multipole expansion and it should be stressed, the the $\ell=0,1$ multipoles need a special treatment akin to this discussed in \cite{r2017}. 

Our scheme is not fully gauge invariant (in the sense discussed in \cite{BMMS,GP}). However we solve the perturbative Einstein equations iteratively anyway and thus we do not see such fully gauge invariant formulation neither necessary nor useful. In potential applications it is important to be able to formulate the final answer in gauge invariant (or gauge controlled) terms, and not necessary to keep such fully gauge invariant formulation at each intermediate (perturbation) step. As the \textit{Lorentzian} Lichnerowicz operator (\ref{eq:Delta_L}) contains only the quantities that are gauge invariant in Regge-Wheeler sense, i.e. $\Delta_L \pert{i}{h}_{\mu\nu}$ does not change under order-$j$ gauge transformation generated with $\pert{j}{\zeta}_\mu$ gauge vector, with $j \geq i$, this a legitimate approach. The gauge vectors $\pert{j}{\zeta}_\mu$ with $j<i$ (entering at order $i$ only the source terms $\pert{i}{S}_{\mu\nu}$) have been already fixed at lower orders $j<i$ to meet the desired asymptotic conditions for metric perturbations (see \cite{r2017} for more details). Moreover, in this approach the sources $\pert{i}{\tilde{S}}^{\Psc}_{\ell}$ are well behaved asymptotically (i.e. their asymptotic behaviour is compatible with the asymptotic behaviour of solutions of homogeneous part of of the scalar wave equation (\ref{eq:scalar_wave})), and there is no need for tedious and obscure procedure of regularizing these sources (see eq.(43) and the comments bellow eq.(88) in \cite{GP}). 

We finish with one \textit{technological} remark. Even though the presented scheme is conceptually very simple, its technical realization, in particular establishing the correct form of source functions, cf. (\ref{eq:alpha}-\ref{eq:gamma},\ref{eq:alpha_comoving}-\ref{eq:beta_comoving}) can be quite involved, and rather unthinkable to achieve in \textit{pre-} computer-algebra era. 

%%%%%%%%%%%%%%%%%%%%%%%%%%%%%%%%%%%%%%%%%%%%%%%%%%%%%%%%%%%%%%%%%%%%%%%%%%%%%%
%%%%%%%%%%%%%%%%%%%%%%%%%%%%%%%%%%%%%%%%%%%%%%%%%%%%%%%%%%%%%%%%%%%%%%%%%%%%%%
%%%%%%%%%%%%%%%%%%%%%%%%%%%%%%%%%%%%%%%%%%%%%%%%%%%%%%%%%%%%%%%%%%%%%%%%%%%%%%

%%%%%%%%%%%%%%%%%%%%%%%%%%%%%%%%%%%%%%%%%%%%%%%%%%%%%%%%%%%%%%%%%%%%%%%%%%%%%%
%%%%%%%%%%%%%%%%%%%%%%%%%%%%%%%%%%%%%%%%%%%%%%%%%%%%%%%%%%%%%%%%%%%%%%%%%%%%%%
%%%%%%%%%%%%%%%%%%%%%%%%%%%%%%%%%%%%%%%%%%%%%%%%%%%%%%%%%%%%%%%%%%%%%%%%%%%%%%
\noindent
\subsubsection*{Acknowledgements.} This research was supported in part by the Polish National Science Centre grant no. DEC-2012/06/A/ST2/00397. 

I acknowledge deeply and with great respect Wolfram \textit{Mathematica} project for providing such a wonderful tool as \textit{Mathematica} for doing research in mathematical physics.

I acknowledge the six months \textit{scientific associate} contract at CERN TH department, January-June 2016, where I had time to develop the ideas that ultimately resulted in this work.
% and I am particularly grateful to Luis \'Alvarez-Gaum\'e for providing propitious atmosphere for those research. 

I am indebt to Piotr Bizo\'n for careful reading and many valuable remarks on the early version of the manuscript.
%I dedicate this work to Piotr Bizo\'n, my mentor for the last fourteen years. It was an honour to become \mbox{Piotr's} collaborator in 2003 and a great pleasure to work with him on all our joint project. I am very grateful for his steady, peaceful support, and many invaluable discussions. No doubt, without Piotr's guidance I would have never come to study nonlinear phenomena in gravitational physics. 

%\end{widetext}

\end{document}